\documentclass[twocolumn,prl, showpacs, superscriptaddress]{revtex4}
\usepackage{amssymb}
\usepackage{amsmath}
\usepackage{bbm}
\usepackage{graphicx}

\begin{document}

\author{Philipp Werner}
\affiliation{Department of Physics, University of Fribourg, 1700 Fribourg, Switzerland}
\author{Martin Eckstein}
\affiliation{Max Planck Research Department for Structural Dynamics, University of Hamburg-CFEL, Hamburg, Germany}
\title{Field-induced polaron formation in the Holstein-Hubbard model}

\date{\today}

\hyphenation{}

\begin{abstract}
We study the effect of strong DC and pulsed electric fields on a Mott insulating system with coupling to optical phonons. A DC field of the order of the gap induces a metallic state characterized by 
polaronic features in the gap region and a partially inverted population. In this quasi-steady state, the field-induced doublon-hole production is balanced by a phonon-enhanced doublon-hole recombination. The photo-excitation of carriers by a pulsed field leads to similar modifications of the electronic structure in the gap region, and an associated reduction of the doublon life-time. We demonstrate that the field-induced localization of electrons effectively enhances the phonon coupling, an effect which should be measureable with time-resolved photoemission spectroscopy.    
\end{abstract}


\maketitle

Strong electric fields and other external perturbations can entirely change the properties of correlated electron materials. 
If a small DC electric field is applied to a Mott insulator, it merely induces a polarization, 
but if the field exceeds a certain threshold, carriers can tunnel across the gap \cite{Taguchi2000,Oka2010, Eckstein2010,Lenarcic2012}, 
which leads to a nonequilibrium metallic state. For an isolated system,  the field-induced carriers quickly heat up to infinite temperature and the current 
vanishes \cite{Mierzejewski2010,Eckstein2011Bloch,Eckstein2013,Fotso2013}, 
but with dissipation, a nonequilibrium steady state with a nonthermal energy distribution may be established. Alternatively, 
mobile charge carriers can be  introduced into a Mott insulator by a laser pulse whose frequency is tuned to the Mott gap \cite{Iwai2003, Okamoto2007, Okamoto2010}. 
Here an efficient dissipation mechanism may cool photo-excited hot carriers before they 
recombine, and thus lead to a photodoped metallic state which is different from a chemically 
doped metal \cite{Eckstein2013photodoping}. 

In real materials, it is the lattice which acts as a heat bath for the electrons. In a metal, the energy transfer from the electrons to the lattice can often be described by 
weak-coupling theory \cite{Sentef2013}, or by the phenomenological two-temperature model \cite{Allen1987}, 
which relies on the assumption that the electron dynamics is fast compared to the phonons.
However, in strongly interacting systems, the time-scales for electronic processes and the lattice dynamics are not 
in general very different. For example, the strongly coupled molecular vibrations in organic Mott insulators have 
frequencies comparable or even larger than the electronic bandwidth \cite{Kaiser2014,Mitrano2013}, and 
conversely, the relaxation times of charge excitations in Mott insulators can become much longer than the 
fast electronic hopping times \cite{Okamoto2007,Mitrano2013,Eckstein2011pump,Lenarcic2013}.  In a 
theoretical description of a dielectric breakdown or photodoping process, it may thus be important to treat 
the electronic system and the lattice on equal footing. 

For a single electron coupled to the lattice, calculations based on exact diagonalization for the Holstein 
model predict intriguing nonlinear transport phenomena \cite{Vidmar2011}, and a relaxation of photo-excited 
carriers on the timescale of the hopping \cite{Golez2012}.
In interacting electron systems, {\it e.g.} the half-filled Mott insulator, exact diagonalization 
\cite{Vidmar2011b,Filippis2012} and density-matrix renormalization group \cite{Matsueda2012}
calculations  
for the Holstein-Hubbard and Holstein-$t$-$J$ models have been used to study the relaxation 
dynamics of photo-carriers. 
For higher excitation densities, the situation is however unclear. 
Understanding the properties of field-induced metallic states, for which nontrivial effects can 
be anticipated due to the competition between the Coulomb repulsion and the electron-phonon 
coupling, remains a challenging problem.

In this work we consider the Holstein-Hubbard model in the Mott insulating phase and drive it out of equilibrium 
by applying strong electric fields. The model describes electrons with local interactions and a coupling to dispersionless phonons, 
\begin{align}
&H(t)=-\sum_{i,\delta,\sigma}v_{i,i+\delta}(t) c^\dagger_{i+\delta,\sigma}c_{i,\sigma}+\sum_i [\omega_0b^\dagger_i b_i + Un_{i,\uparrow}n_{i,\downarrow}]\nonumber\\
&+\sum_i [-\mu(n_{i,\uparrow}+n_{i,\downarrow})+g(n_{i,\uparrow}+n_{i,\downarrow}-1) (b^\dagger_i+b_i)].
\label{H}
\end{align}
Here, $U$ is the on-site repulsion, $\mu$ the chemical potential of the electrons with creation operators 
$c^\dagger_\sigma$ and density operators $n_\sigma$, $b^\dagger$ the creation operator for Einstein phonons of frequency $\omega_0$, and $g$ the electron-phonon coupling strength. In a gauge with pure vector potential $A(t)$, the electric field $E(t)=-\partial_t A(t)$ enters via a
time-dependent phase for the hopping terms, $v_{i,j}(t)=v_{i,j}^0\exp[i(R_i-R_j)A(t)]$. We will consider an infinite-dimensional hypercubic lattice and apply the field 
along the body-diagonal \cite{Turkowski2005}.  
The density of states is $\rho(\epsilon)= \exp(-\epsilon^2/W^2)/\sqrt{\pi}W$, and we measure energy 
(time) in units of $W$ ($1/W$).
Throughout the paper we focus on Mott insulating systems ($U>$ bandwidth) 
which are strongly coupled to optical phonons with frequencies comparable to the bandwidth ($\omega_0=1$). 
Such a regime may be realized in some organic Mott insulators \cite{Okamoto2007,Kaiser2014}, where the electrons
couple to molecular vibrations. 

An approximate treatment of the Holstein-Hubbard model 
is possible within the framework of nonequilibrium dynamical mean field theory (DMFT) 
\cite{Freericks2006, Aoki2013, Werner2013_phonon}. The precise formalism 
has been detailed in Refs.~\cite{Werner2013_phonon, Werner2007_phonon}.  
Electric fields can be incorporated via the DMFT self-consistency equations 
in the same manner as without phonons \cite{Aoki2013}. 
To solve the DMFT equations, we use a strong-coupling impurity solver based 
on the non-crossing approximation \cite{Eckstein2010nca, Werner2013_phonon}. This solver has been 
shown to capture the competition between on-site repulsion and phonon-mediated attraction, and is 
expected to give qualitatively correct results if applied to the Mott phase.

\begin{figure}[t]
\begin{center}
\includegraphics[angle=-90, width=0.49\columnwidth]{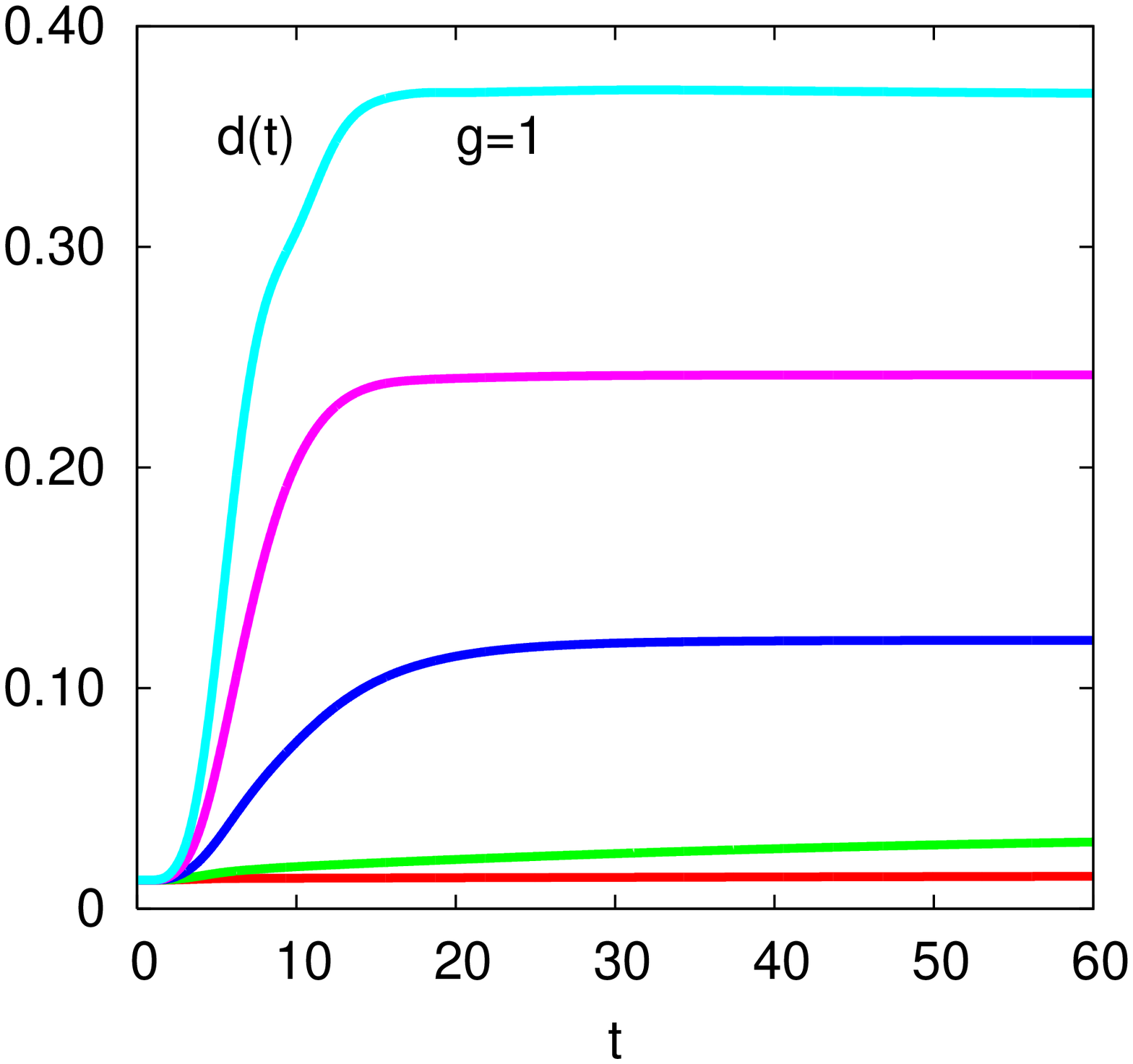}
\includegraphics[angle=-90, width=0.49\columnwidth]{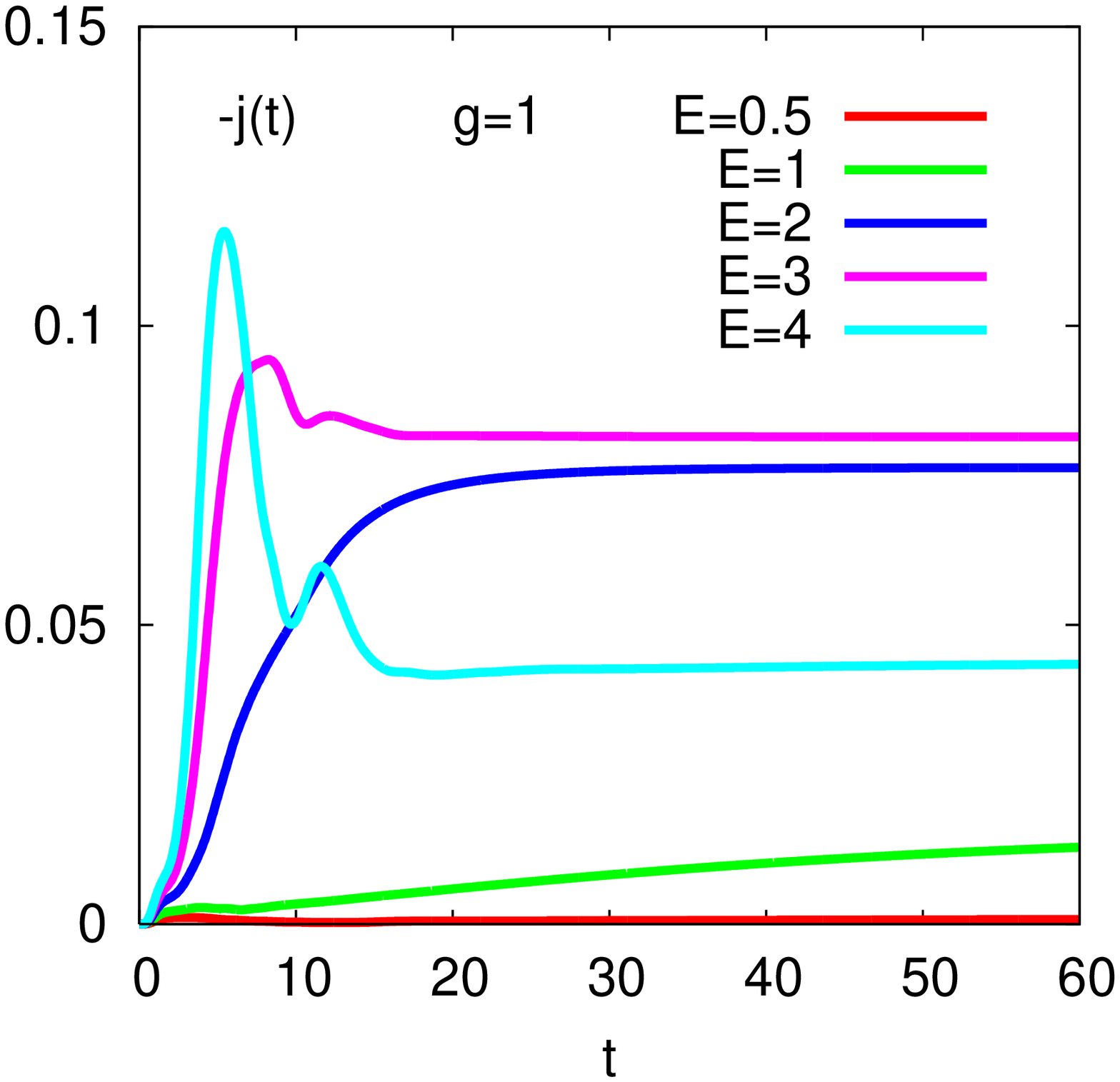}
\caption{
Time-evolution of the double occupancy $d$ and current $j$ for different static fields $E$. 
The parameters are $U=5$, $g=1$ and the inverse temperature of the initial state is $\beta=5$.  
}
\label{fig_current}
\end{center}
\end{figure}   

{\it Static fields.} We first study the dielectric breakdown in a Mott insulating Holstein-Hubbard model with 
$U=5$ and $g=1$. To reduce transients, we smoothly switch on the electric field to the value $E$ within a time $t_\text{switch}\approx 8$. 
Figure~\ref{fig_current} plots the time-evolution of the double occupancy $d(t)=\langle n_\uparrow(t)n_\downarrow(t)\rangle$ 
and the current $j(t)\equiv j_\sigma=\langle\frac{1}{L}\sum_k\partial_k\epsilon_{k+A(t)}n_{k\sigma}(t)\rangle$ for 
different values of $E$. 
At short times and for weak fields, one observes a steady increase in $d(t)$, which reflects the field-induced 
increase of the carrier density. Because these carriers are cooled by phonon-scattering, the current $j(t)$ 
increases roughly linearly with the number of carriers ({\it i.e.} here the phonons act essentially as a heat bath 
\cite{Eckstein2013}). For $E \gtrsim 1.5$ 
a dielectric breakdown occurs, which is marked by a rapid increase in the 
current and number of doublons. Despite the strong field, the nonequilibrium Holstein-Hubbard model 
quickly settles into a quasi-steady state characterized by a large doublon and hole density, and by a large 
current. In this state, energy is continuously flowing to the lattice.
For the largest currents in Fig.~\ref{fig_current} about $4$ phonons per site are excited by $t=60$.
(In a real material, anharmonic effects will eventually 
become important, so that the Hubbard-Holstein description is appropriate only for sufficiently short time.)

\begin{figure}[t]
\begin{center}
\includegraphics[angle=-90, width=0.49\columnwidth]{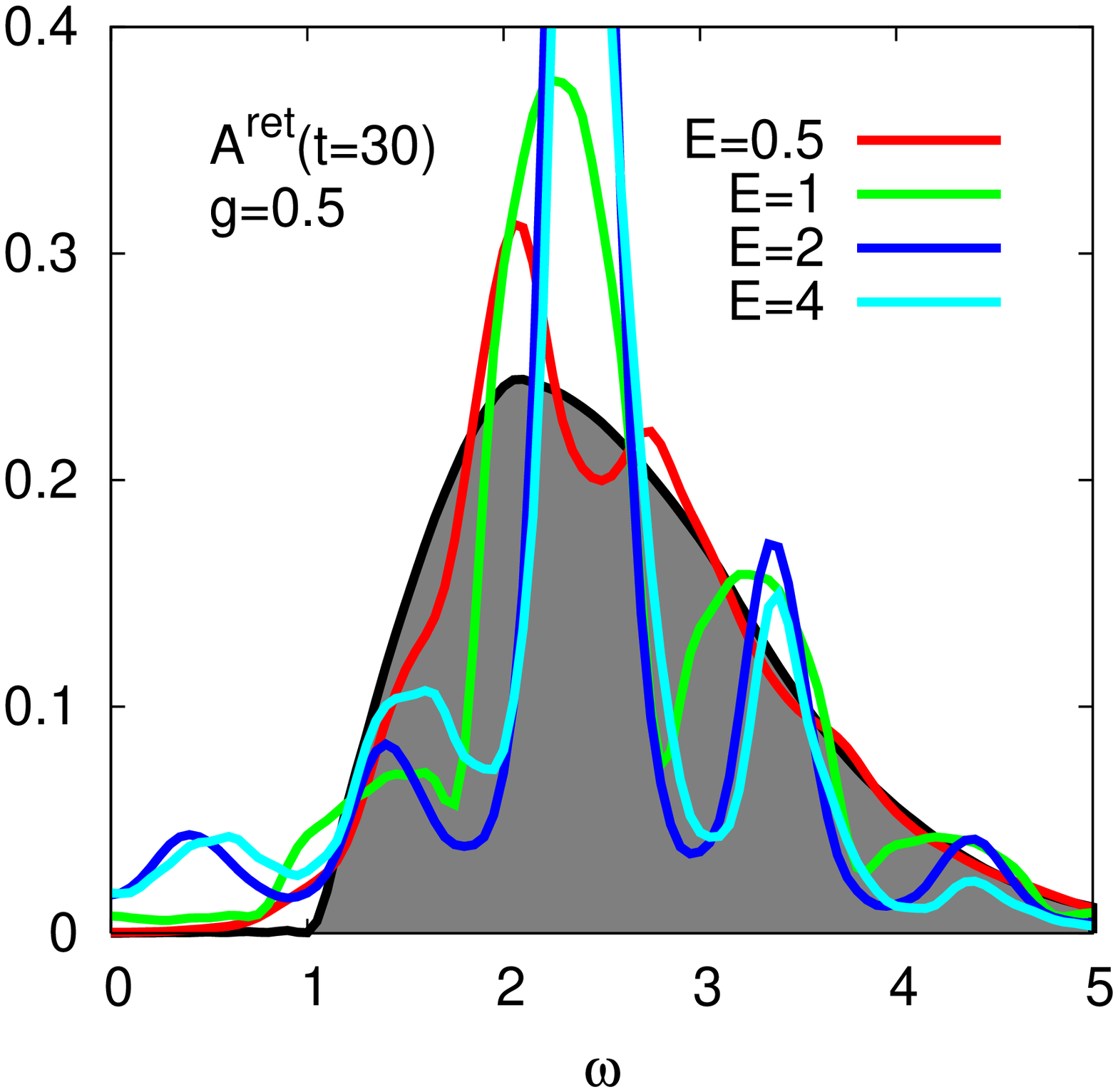}
\includegraphics[angle=-90, width=0.49\columnwidth]{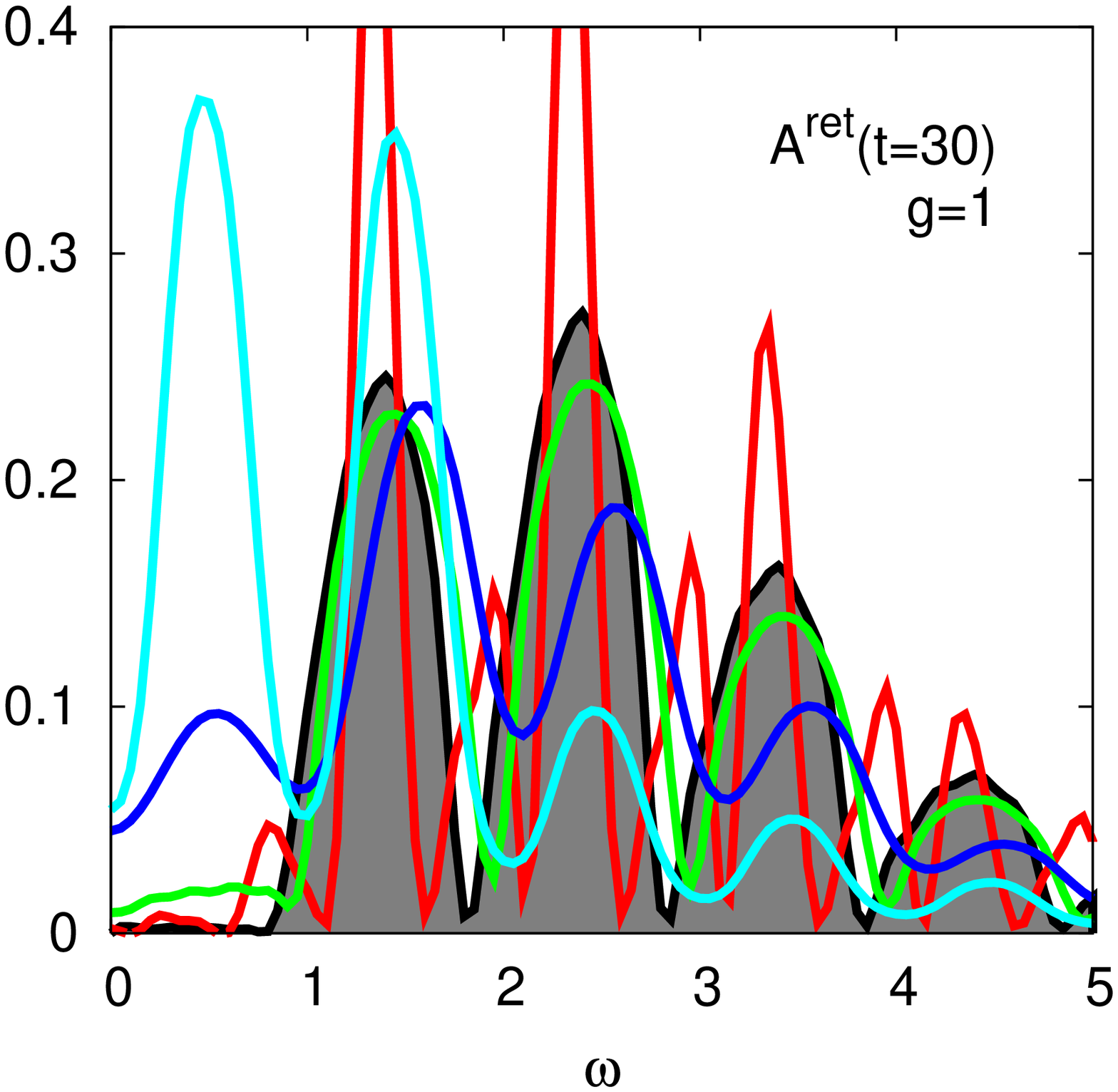}\\
\hspace{4mm}\includegraphics[angle=-90, width=0.46\columnwidth]{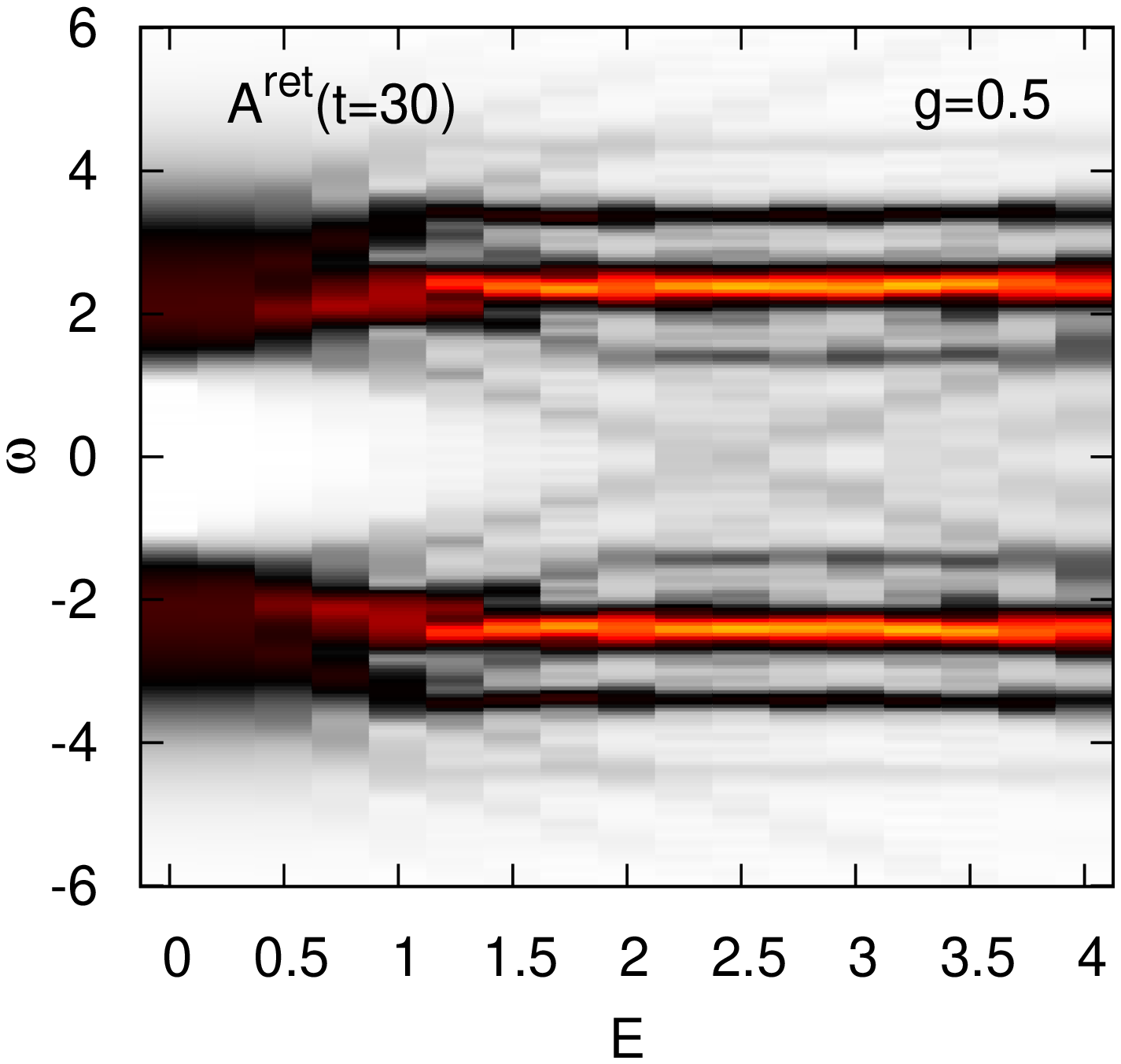}\hfill
\includegraphics[angle=-90, width=0.46\columnwidth]{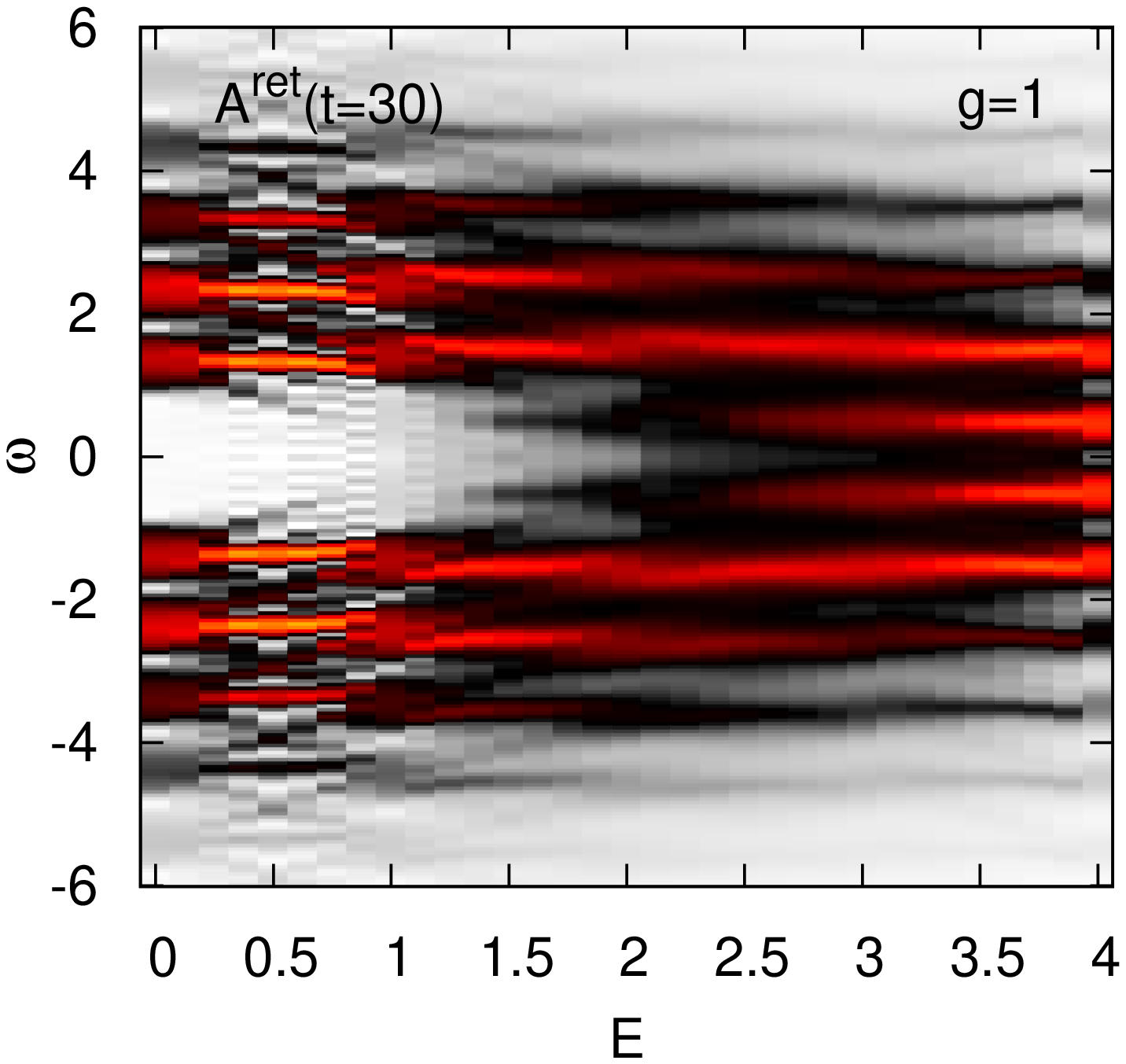}\\
\vspace{1.5mm}\mbox{}\\
\includegraphics[angle=-90, width=0.49\columnwidth]{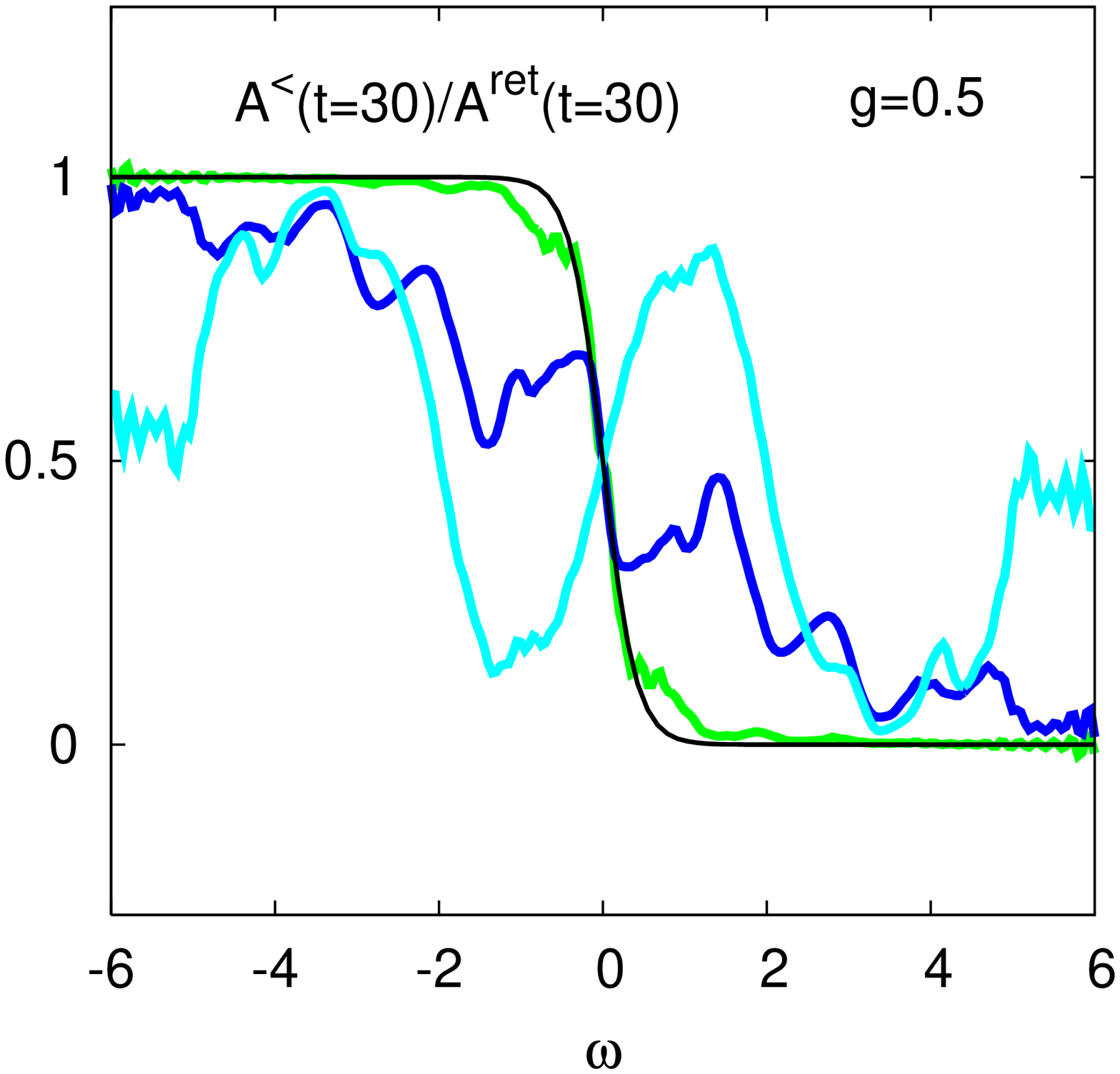}
\includegraphics[angle=-90, width=0.49\columnwidth]{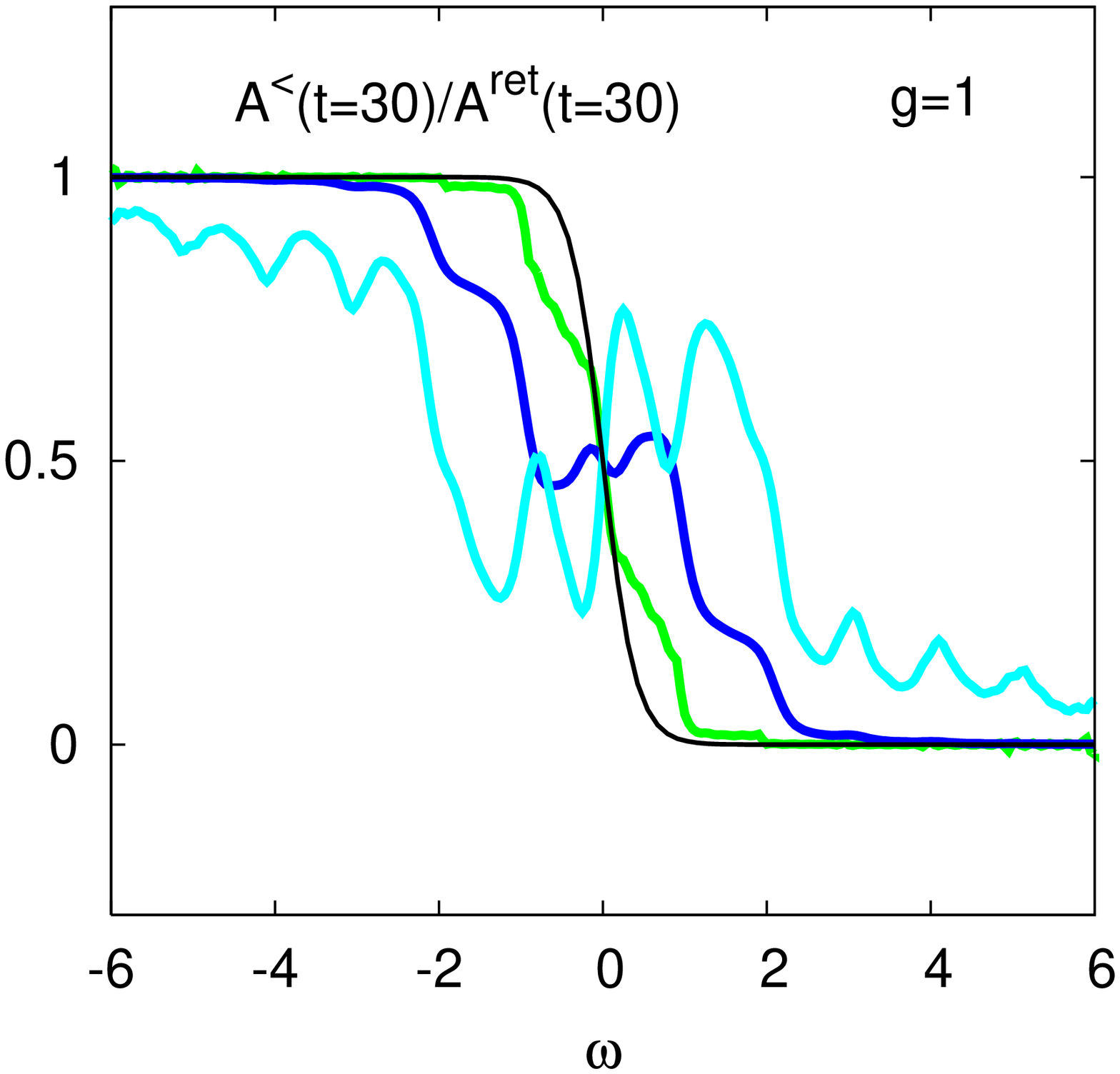}
\caption{Top four panels: Spectral function $A^\text{ret}(\omega,t=30)$ for $g=0.5$ (left) and $g=1$ (right), 
plotted for different values of $E$ ($U=5$). 
The shaded area in the upper panels indicates the spectrum for $E=0$.
Lower panels: ``distribution function" $A^</A^\text{ret}$ of the quasi-steady state. 
In black we show the equilibrium distribution functions for $\beta=5$.
}
\label{fig_spectra_E}
\end{center}
\end{figure}   

\begin{figure}[t]
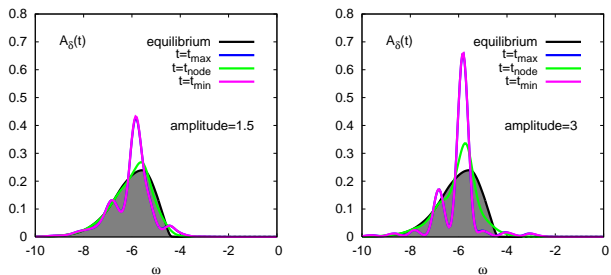

\begin{center}
\includegraphics[angle=-90, width=0.49\columnwidth]{figure3a.eps}
\includegraphics[angle=-90, width=0.49\columnwidth]{figure3b.eps}
\caption{Time-resolved photoemission spectrum $A_\delta(\omega,t)$ for probe pulse width $\delta=4$ in a system driven by a slow $\Omega_\text{pulse}=0.0833$ single-cycle pulse with amplitude much smaller than the gap ($U=12$, $g=0.5$). The measurements are taken at the maximum of the first half-cycle ($t=t_\text{max}=21.8$), at the node ($t=t_\text{node}=31.4$) and at the minimum of the second half-cycle ($t=t_\text{min}=41.1$).
}
\label{fig_THz}
\end{center}
\end{figure}   

Further insights into the quasi-steady state can be obtained from the nonequilibrium spectral functions 
$A^{\text{ret}, <}(t)=-\frac{1}{\pi}\int_t^\infty dt'e^{i\omega (t'-t)}\text{Im}G^{\text{ret}, <}(t, t')$.    
The top four panels in Fig.~\ref{fig_spectra_E}  plot $A^\text{ret}(\omega,t=30)$ for $g=0.5$ and $g=1$. 
The spectrum shows an interplay between field-induced sidebands of the Hubbard bands 
(Wannier-Stark peaks), whose positions depend linearly on the field, 
and phonon sidebands, which are spaced by $\omega_0=1$ and do not shift with the field.  
Interestingly, prominent phonon side-bands appear in the presence of an electric field even in the weakly 
coupled system, where the initial density of states shows no phonon features. 
The field strength $E\gtrsim 1$ needed to induce well-separated phonon bands is approximately the same as the field 
strength which leads to clearly separated Wannier-Stark peaks in the absence of electron-phonon coupling. 
We thus interpret the observed changes in the spectral function as an effective enhancement of the 
electron-phonon coupling due to the field-induced localization of the electrons,
in analogy to the enhancement of Coulomb interaction effects in the Wannier-Stark states in metals
\cite{Freericks2008, Eckstein2011damping}. 

This effect might be observable in Mott insulators such as (ET-F2TCNQ) \cite{Okamoto2007},  
if the material is driven by intense THz pulses (frequency $\ll$ gap). To illustrate this, we plot in 
Fig.~\ref{fig_THz} the time-resolved photoemission spectrum \cite{Freericks2009} 
$A_\delta(\omega, t_p)=\int dt dt' s(t-t_p)s(t'-t_p)e^{i\omega(t-t')}G^<(t,t')$, measured 
with a Gaussian probe pulse envelope $s(t)=\exp(-t^2/\delta^2)$. We consider a large gap insulator with 
relatively weak phonon coupling ($U=12$, $g=0.5$), which is driven by a single-cycle field pulse 
with frequency $\Omega_\text{pulse}=0.0833\ll \text{gap}$. 
The curves in the figure show the photoemission spectrum measured near the maximum 
($t=t_\text{max}$), the minimum ($t=t_\text{min}$), and the node ($t=t_\text{node}$) of the pump pulse. 
The width of the probe-pulse ($\delta=4$) is chosen such that the field-induced phonon sidebands can 
be resolved in frequency, while the spectra measured at $t=t_\text{max}, t_\text{min}$ and 
$t_\text{node}$ can be clearly separated in time. 
Already for a pump pulse amplitude of $1.5$ one can clearly identify the field-induced phonon side-bands in the spectra 
measured near the maxima of the pump pulse, while the spectrum near the node differs only little from the featureless 
equilibrium result. At the larger field amplitude of $3$, the phonon features are even more prominently visible. In both cases, 
the pump pulse produces only a negligible number of doublons, $O(10^{-6})$, so that the spectra measured at the first 
maximum and the first minimum are almost identical.   

In the strong-coupling regime ($g=1$, right panels of Fig.~\ref{fig_spectra_E}) the Hubbard bands are split into phonon side-bands
already in the thermal initial state. Here, one observes Wannier-Stark side-bands emanating from the phonon peaks already at weak fields. 
The Wannier-Stark bands intersect at fields  $E\approx n\omega_0+0.5$ ($n=0,1,\ldots$), leading to much 
sharper phonon side-bands due to field-induced localization, while the phonon bands broaden around the 
resonances $E\approx n\omega_0$ ($n=1,2,\ldots$). 
For field strengths larger than the gap size ($E\gtrsim 1.5$) one notices the appearance of states in the gap region. 
These in-gap states correspond to polarons (doublons or holes dressed by phonons) and are reminiscent of the
side-bands appearing in the equilibrium spectral functions of doped Mott insulators  
(see Supplementary Material).
The in-gap states on the $\omega>0$ side are highly occupied and correspond 
to dressed doublons, while the in-gap states on the $\omega<0$ side are depleted and correspond to dressed holes. 
We illustrate this partial inversion of the occupation in the gap region by plotting the 
``nonequilibrium distribution function"  $A^<(\omega, t=30)/A^\text{ret}(\omega, t=30)$  
in Fig.~\ref{fig_spectra_E} (bottom panels).
 
\begin{figure}[t]
\begin{center}
\includegraphics[angle=-90, width=0.48\columnwidth]{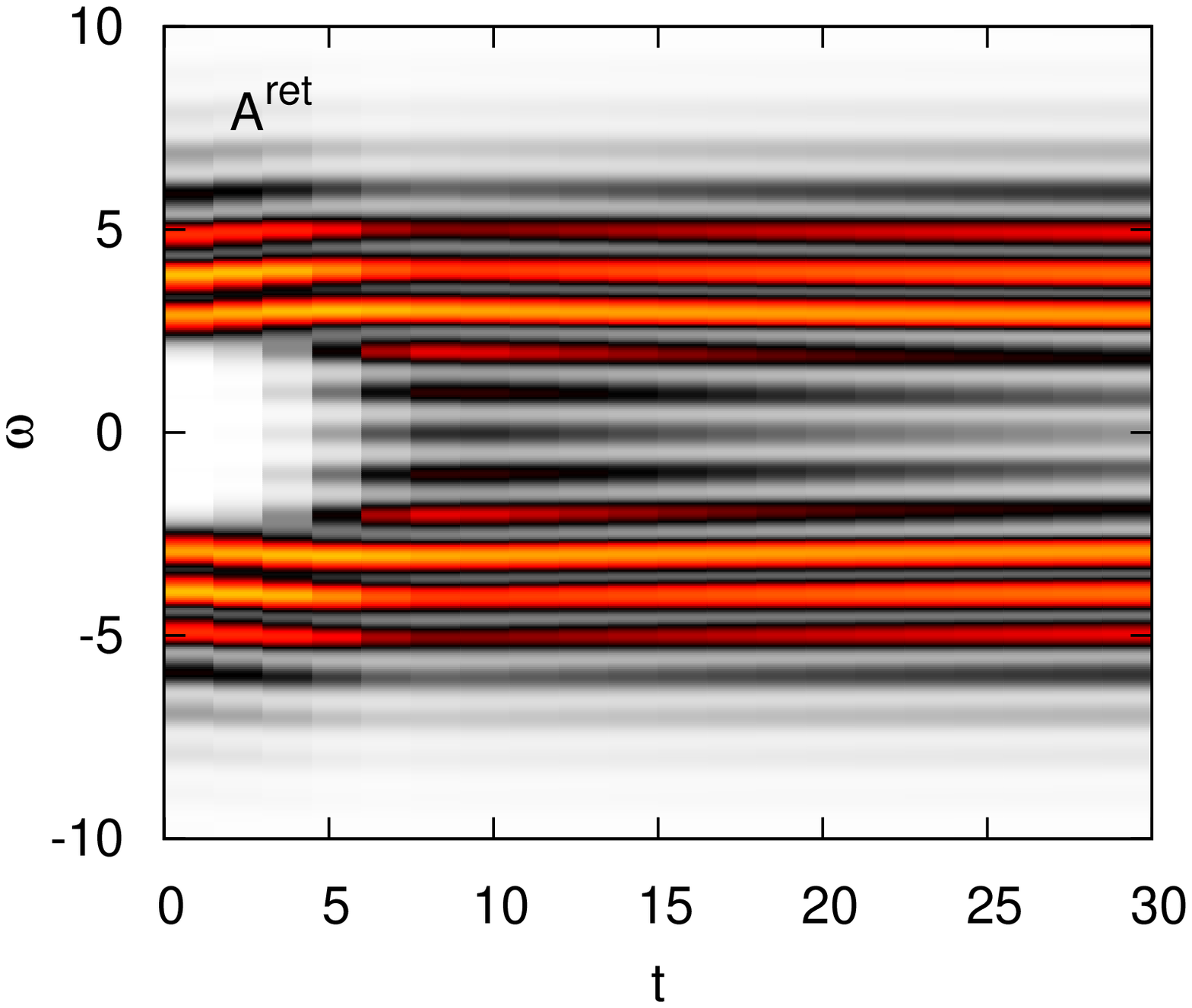}
\includegraphics[angle=-90, width=0.48\columnwidth]{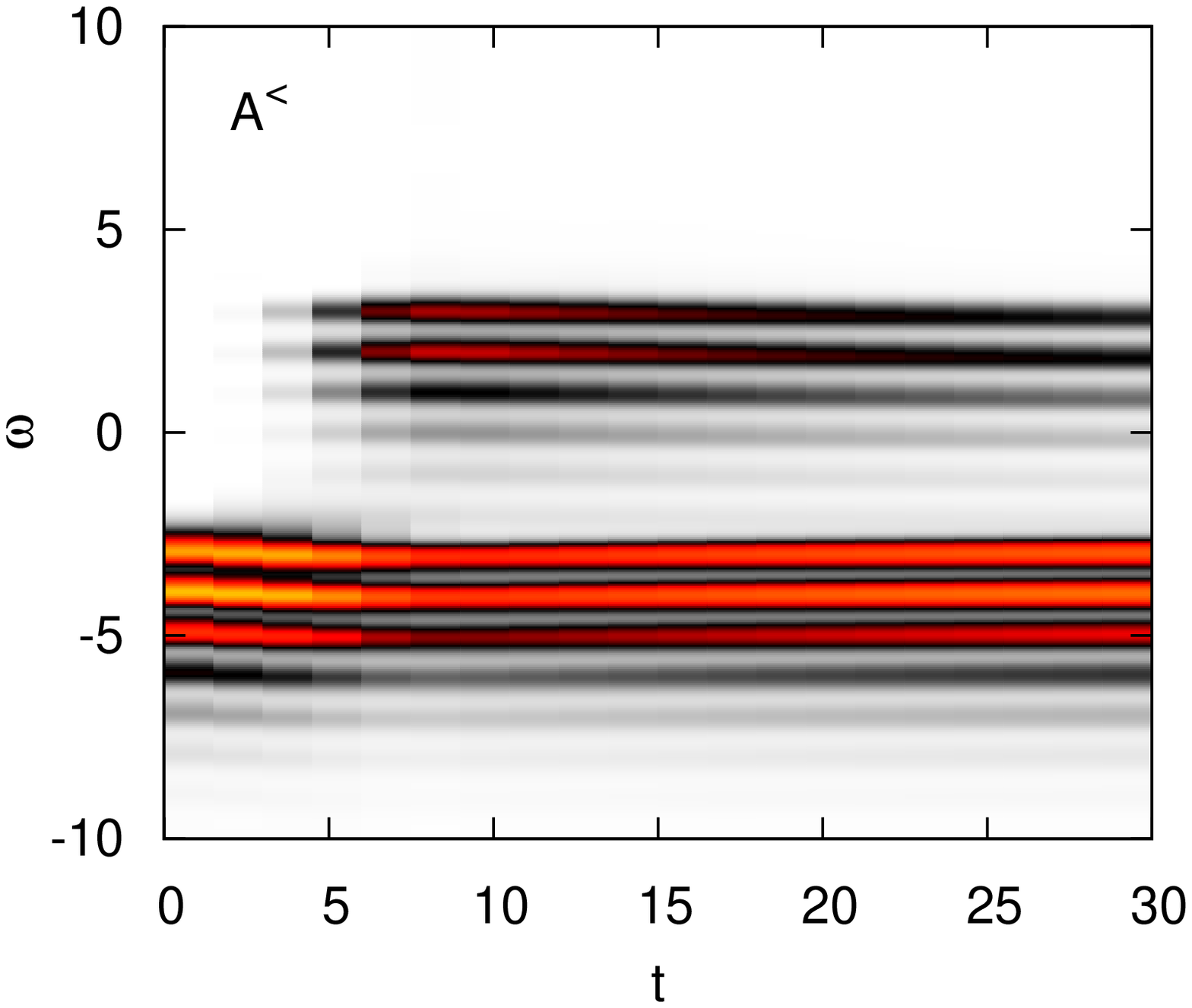}\\
\vspace{5mm}
\includegraphics[angle=-90, width=0.48\columnwidth]{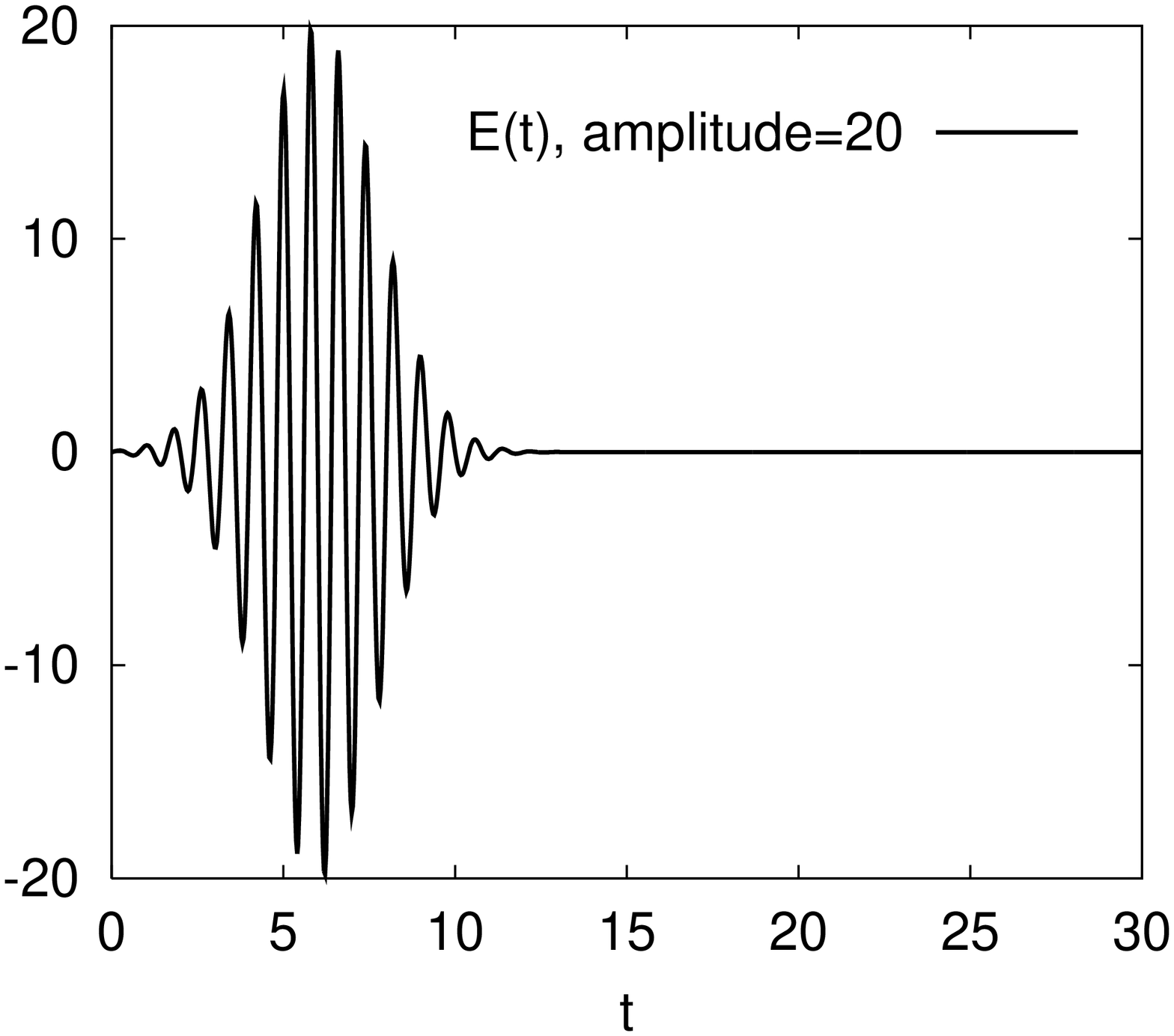}
\includegraphics[angle=-90, width=0.48\columnwidth]{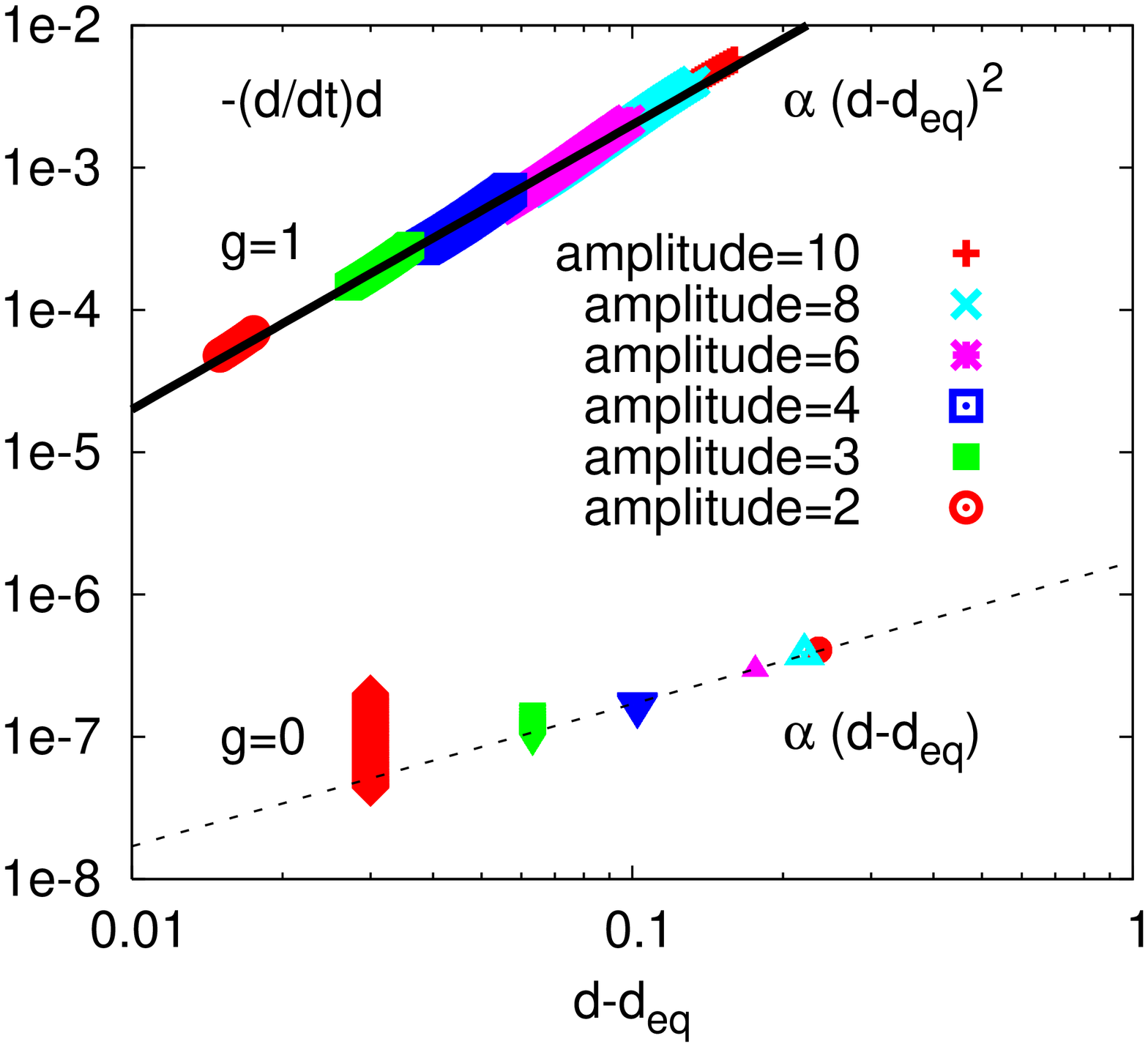}
\caption{Top panels: Time-resolved spectral function $A^\text{ret}(\omega,t)$ (left) and $A^<(\omega,t)$ (right) after a pulse excitation ($U=8, g=1$). 
Bottom left: Few-cycle pulse with $\Omega_\text{pulse}\approx 8$. Bottom right: Doublon decay-rate as a function of $d-d_\text{eq}$ for different pulse amplitudes (photo-doping concentrations).
}
\label{fig_spectra_doping}
\end{center}
\end{figure}

{\it Photodoping. } 
Similar nonthermal spectral functions with polaronic mid-gap states appear after photo-doping. 
We first analyze this effect for a Mott insulator with $U=8$ and a few-cycle electric field pulse with frequency $\Omega_\text{pulse}\approx 8$. 
This pulse excites electrons across the gap and 
creates a population in the middle of the upper Hubbard band.  
In the absence of electron-phonon coupling the occupied part of the spectral function relaxes very slowly \cite{Eckstein2011pump} 
and there is little change in the total spectral function. In the Holstein-Hubbard model, the photo-doping leads to 
modifications in the spectral function, which depend strongly on the concentration of carriers and on the phonon 
coupling strength (Fig.~\ref{fig_spectra_doping}). For $g=1$, polaronic in-gap states appear, which are populated 
by doublons on the $\omega>0$ side and by holes on the $\omega<0$ side. 

The formation of polaronic states has profound effects both on the relaxation and on the recombination of photo-excited carriers. In the absence of electron-phonon coupling, the life-time of photo-doped carriers is essentially determined by the gap-size, and it is very slow: $d(t)\sim \exp(-t/\tau)$, with $\tau$ depending exponentially on the gap size \cite{Sensarma2010, Eckstein2011pump}. In the model with strong phonon-coupling, the doping-induced in-gap states lead to a doping-dependent doublon-hole life-time. If we plot the recombination rate $(-{\rm d}/{\rm d} t)d$ as a function of doublon concentration $d-d_\text{eq}$ \cite{footnote_deq} for different pulse amplitudes (doping concentrations), we find that the rate grows with the concentration approximately like $(d-d_\text{eq})^{2}$, see bottom right panel of Fig.~\ref{fig_spectra_doping}. This quadratic dependence indicates that the rate is determined by the probability of electron-like polarons meeting hole-like polarons. The relaxation mechanism in the presence of phonons is thus entirely different from the kinetically blocked, exponentially slow decay in the Hubbard model. (The decay rate for $g=0$ scales linearly with $d-d_\text{eq}$.)

\begin{figure}[t]
\begin{center}
\includegraphics[angle=-90, width=0.48\columnwidth]{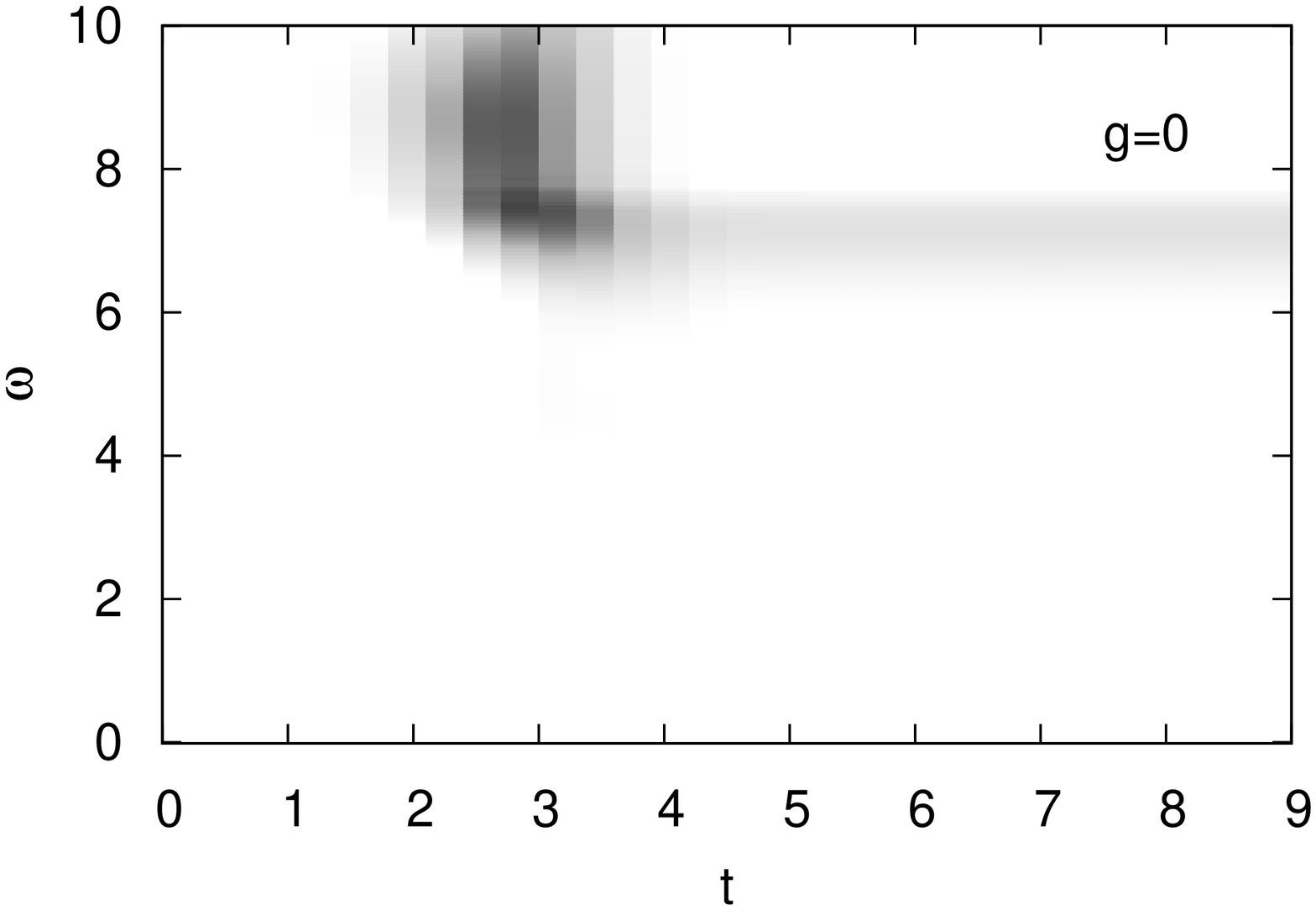}
\includegraphics[angle=-90, width=0.48\columnwidth]{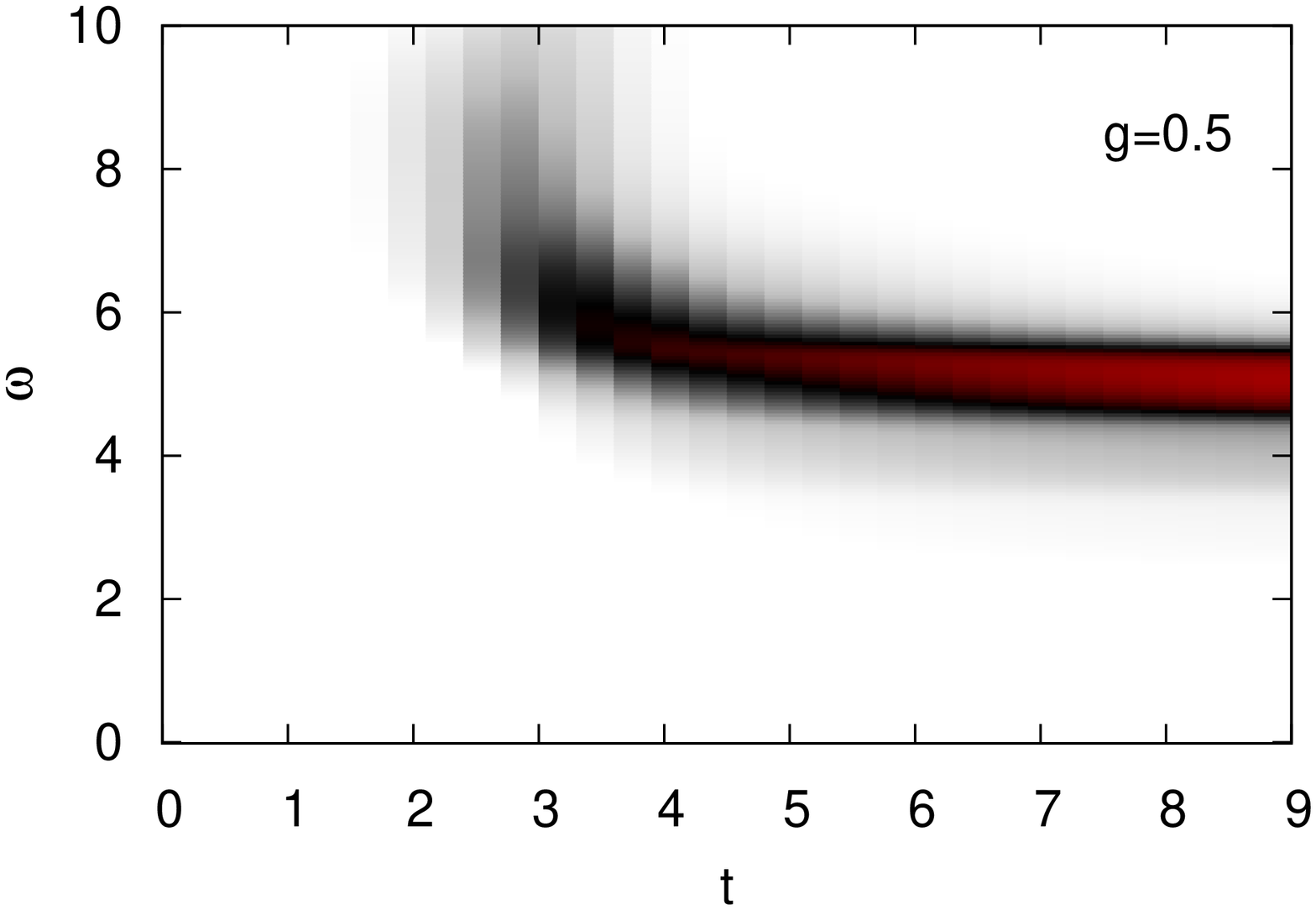}\\
\vspace{2mm}
\includegraphics[angle=-90, width=0.48\columnwidth]{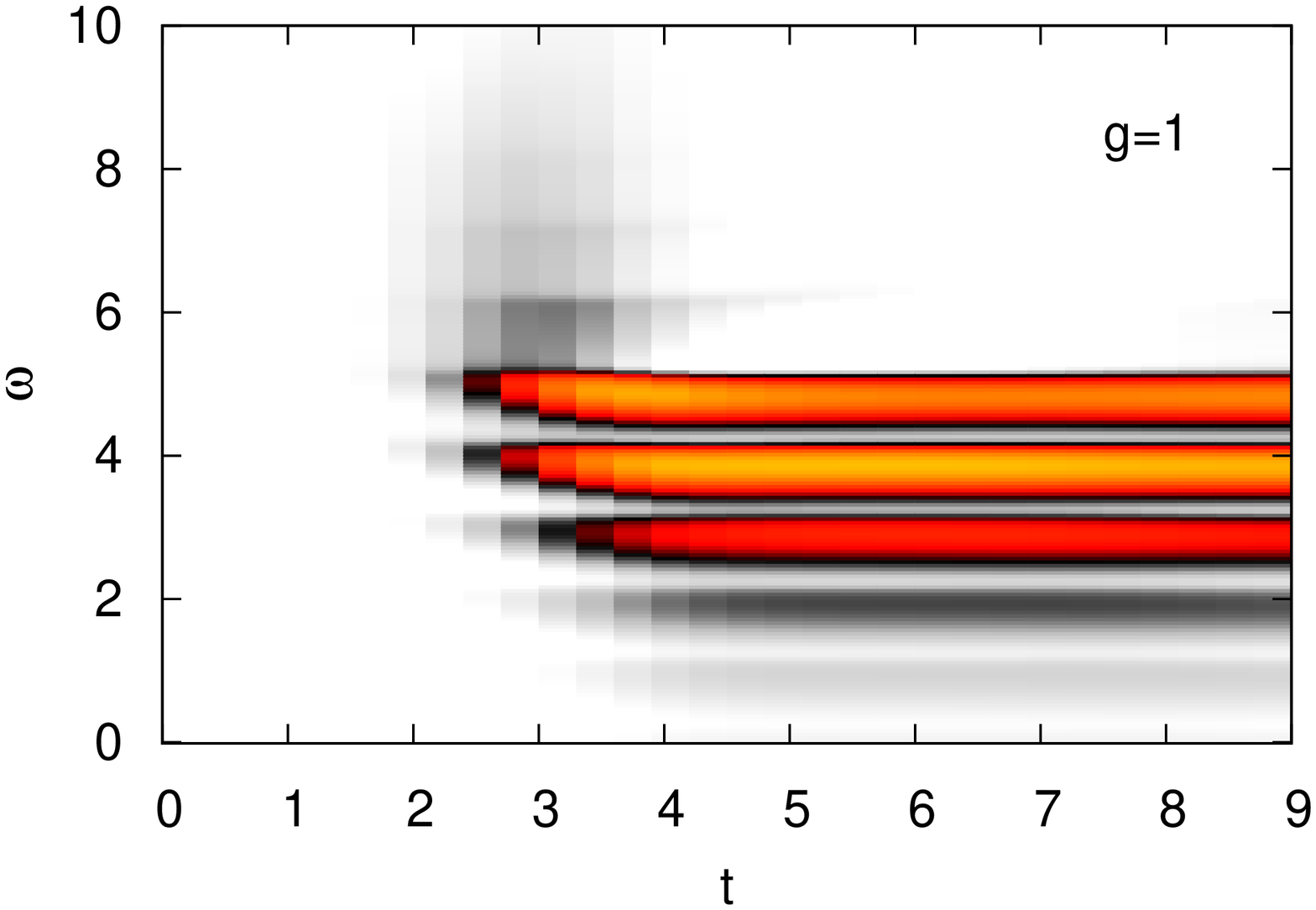}
\includegraphics[angle=-90, width=0.48\columnwidth]{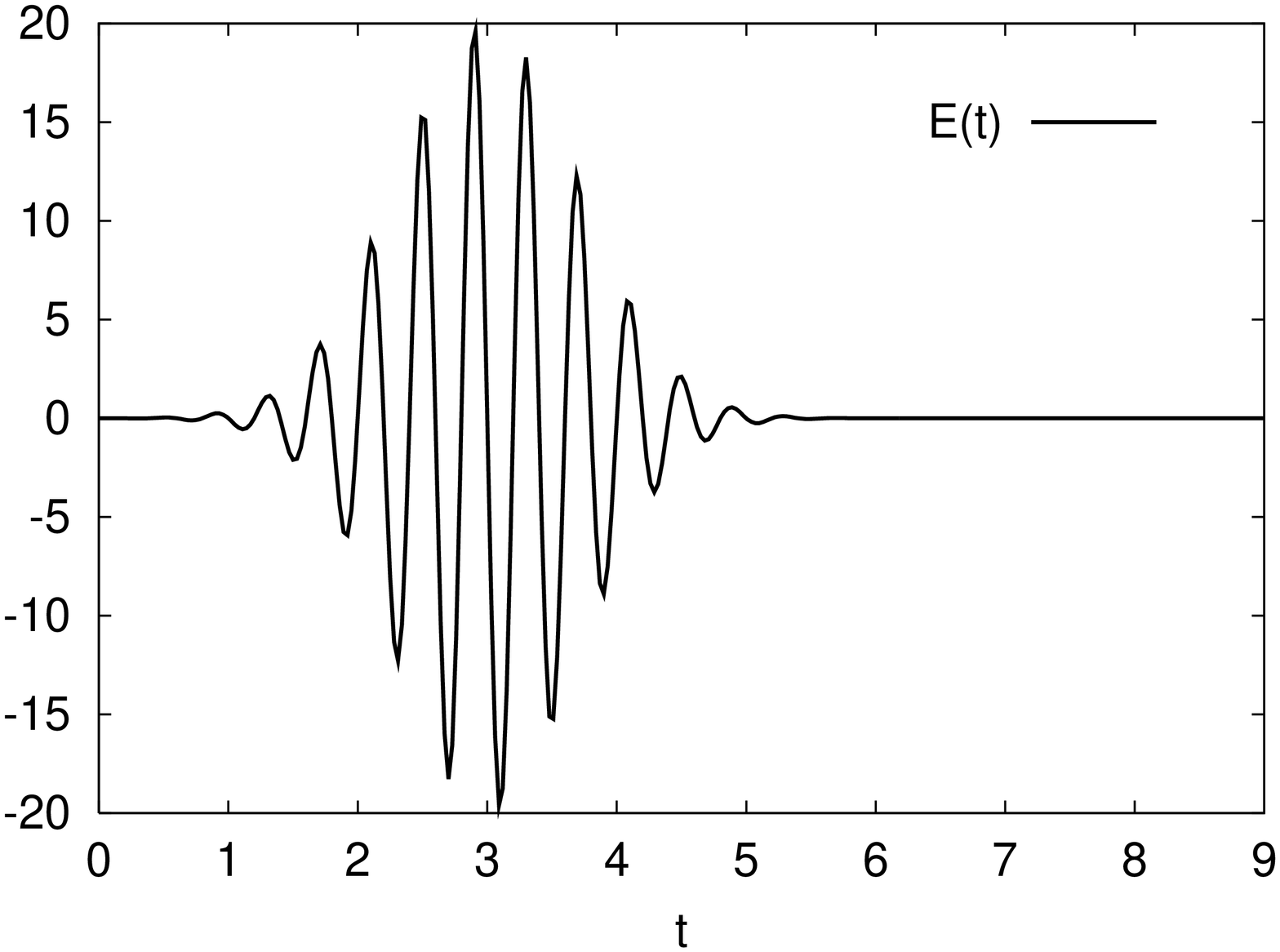}
\caption{Time-resolved spectral function $A^\text{ret}$ during and shortly after a pulse excitation. The interaction is $U=12$ and the pulse frequency is $\Omega_\text{pulse}\approx 15$ (i.e.  
doublons are inserted at the upper edge of the Hubbard band). 
}
\label{fig_cooling}
\end{center}
\end{figure}   

To illustrate the cooling of the photo-doped carriers due to electron-phonon scattering we consider a large-gap insulator ($U=12$) subject 
to an intensive few-cycle pulse with $\Omega_\text{pulse}\approx 15$, which creates a population at the upper edge of the upper band. 
Plots of $A^<(\omega,t)$ for different values of $g$ are shown in Fig.~\ref{fig_cooling}. In the absence of electron-phonon coupling only a small number of doublons is inserted (due to the small overlap of the upper Hubbard band with the power-spectrum of the pulse), and these photo-doped carriers remain confined to the upper edge of the band up to times $t\gg 10$. For 
$g=0.5$, the spectral weight shifts to lower energies and piles up at the lower band edge already at $t \approx 8$.  
For weak coupling, the relaxation (energy dissipation rate) can be estimated from the drift velocity of the spectral weight from the upper band 
edge to the lower band edge. Our simulations indicate that this velocity is determined by the ratio $g^2/\omega_0$ (see Supplementary Material). 
For even larger electron-phonon coupling the Hubbard bands broaden due to pronounced phonon sidebands and the appearance of in-gap states. In the simulation for $g=1$ one observes a rapid population of these side-bands by the cooled polarons. The cooling of photo-induced doublons from an energy $\Omega/2\approx 7.5$ to the lowest phonon-peaks in the upper Hubbard band ($\omega\approx 4$) occurs already during the application of the pulse (before $t=4$), and the subsequent relaxation and polaron-formation (appearance of doping induced in-gap states) is also a very fast process. 

Finally, the spectral weight plots in Fig.~\ref{fig_cooling} also suggest that the number of carriers produced by the high-frequency 
pulse is enhanced by the phonon coupling. We analyzed the $g$-dependence of the doping concentration, and compared it with the result expected 
for a rigid spectral function which does not respond to the pulse. If  the field-strength is in the linear regime, we expect 
a number of doped carriers proportional to the convolution $N_\text{naive}=\int  d\omega' d\omega A^\text{ret}(\omega) 
p(\omega)A^\text{ret}(\omega-\omega')f(\omega-\omega')(1-f(\omega))$, 
where $p(\omega)$ is the powerspectrum of the pulse 
and $f(\omega)$ is the fermi distribution function for $\beta=5$.
While $N_\text{naive}$ increases by a factor $1.7$ if $g$ is increased from $0$ to $1$, due to the broadening of the spectral function, the number of doped carriers in the linear regime increases by a factor of about $4.4$. This must be due to matrix element effects. Outside the linear regime, the absorption in the phonon coupled system becomes much larger than in the Hubbard model: for pulse amplitude $20$ (Fig.~\ref{fig_cooling}), the number of photo-doped carriers is about $17$ times larger for $g=1$ than for $g=0$. 
% [see doping_largefreq.gnu and efieldhighfreq_doping.gnu]

In conclusion, we have shown that field-induced carriers in the Holstein-Hubbard model cool down through the interaction with 
phonons and form polaronic states, which leads to characteristic changes in the time-resolved photoemission spectrum. 
The appearance of dressed electron and hole-like carriers also has a significant effect on the carrier recombination time. 
Even without doublon-hole production, 
electric fields can induce polaronic features because the electron localization in strong fields  
enhances the effect of the phonon coupling.  Our results demonstrate that electron-phonon scattering and dynamical lattice effects can lead to 
a response of the system to external fields which is qualitatively different from that of a Hubbard model, and even of a 
Hubbard model coupled to an equilibrium phonon bath. 
These effects may be observable in time-resolved photo-emission measurements or optical conductivity measurements on organic Mott insulators.

{\it Acknowledgements}. We thank H. Aoki, D. Baeriswyl, D. Golez, A. S. Mishchenko, T. Oka, N. Tsuji and L.~Vidmar for useful discussions. The calculations were run on the UniFr cluster. PW is supported by FP7/ERC starting grant No. 278023.

\newpage

\noindent\\
{\bf Supplementary Material}

In these supplementary notes we present equilibrium spectral functions of the chemically doped Holstein-Hubbard model and discuss the time-resolved optical conductivity after a photo-doping pulse. We also present data supporting the $g^2/\omega_0$-dependence of the energy dissipation rate in phonon-coupled systems.

\noindent\\
{\bf Doped equilibrium model}

\begin{figure}[b]
\begin{center}
\includegraphics[angle=-90, width=0.48\columnwidth]{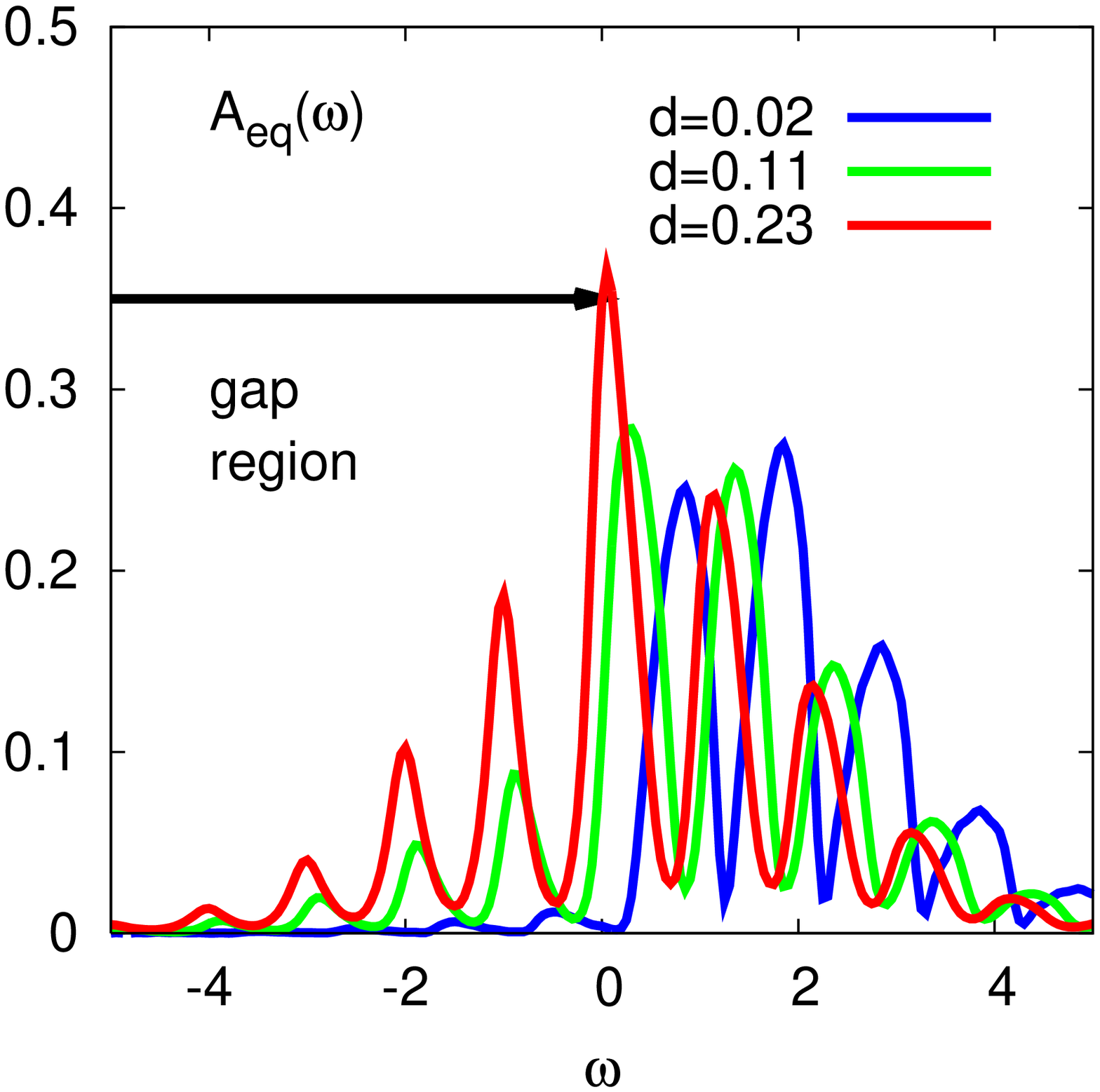}
\includegraphics[angle=-90, width=0.48\columnwidth]{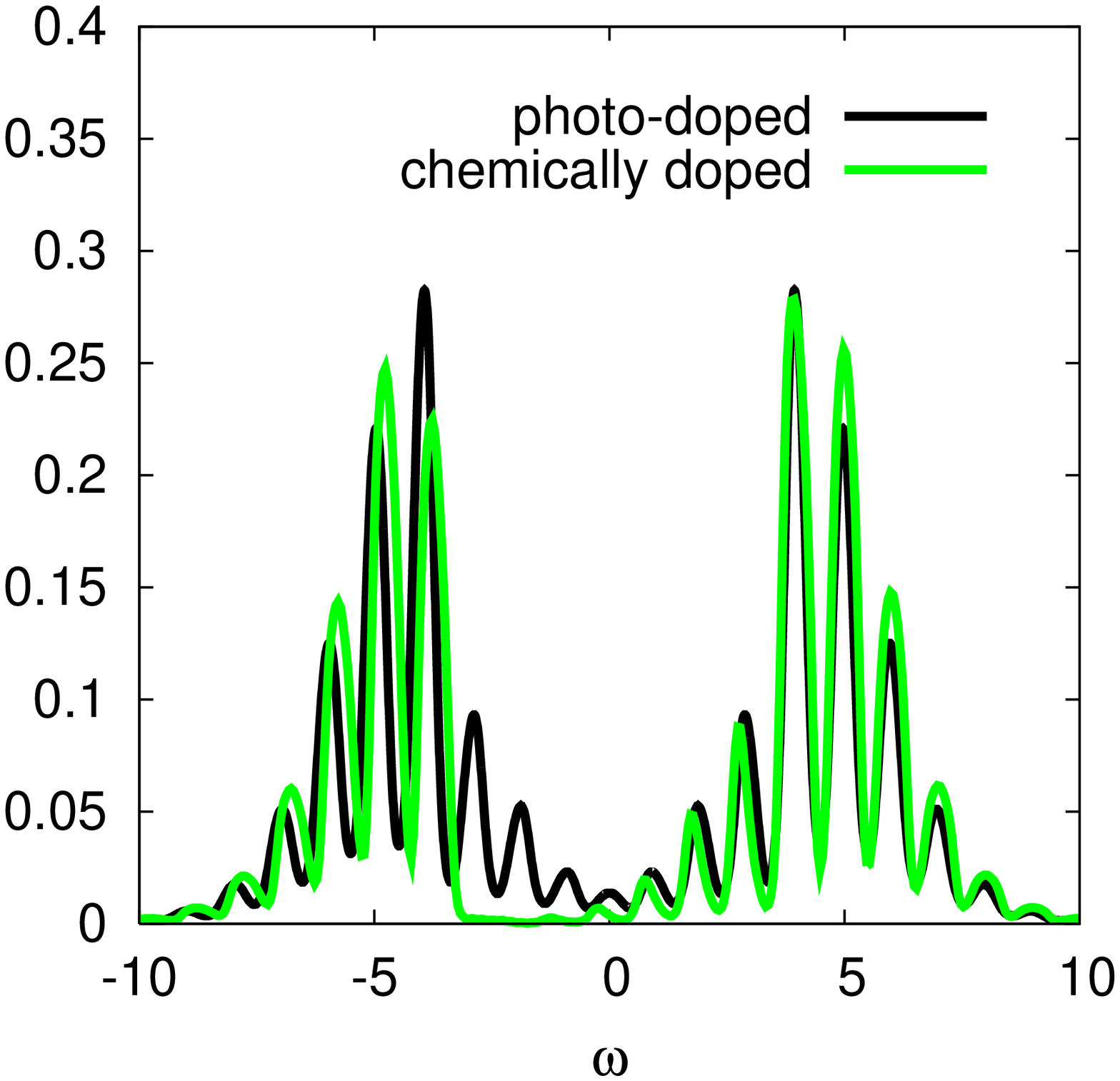}
\caption{Left panel: Equilibrium spectral functions (upper Hubbard band) for the doped Holstein-Hubbard model with $U=10, g=1, \beta=5$. Right panel: Comparison between the nonequilibrium spectral function of a strongly photo-doped 
state ($d=0.11$) and a chemically doped state with a comparable value of $d$. The equilibrium spectral function has been shifted to 
match the position of the dominant peak.
}
\label{fig_equilibrium}
\end{center}
\end{figure}

Polaronic in-gap states similar to the ones observed in the field-induced metallic systems appear in the equilibrium spectral function of the Holstein-Hubbard model away from half-filling (Figure~\ref{fig_equilibrium}). The left panel shows the upper Hubbard band of electron-doped systems with $U=10, g=1, \beta=5$. As the number of doublons increases, more and more prominent side-bands (corresponding to electron-like polarons) appear in the gap region. The spectral features are almost identical to those of a photo-doped system with comparable doublon density (see right panel). This indicates that the cooling by the phonons is effective and that we can view the photo-doped system as a cold (``$\beta=5$") state in the subspace of the Hilbert space corresponding to a certain number of doublons and holes. 

\noindent\\
{\bf Optical conductivity}

\begin{figure}[t]
\begin{center}
\includegraphics[angle=-90, width=0.48\columnwidth]{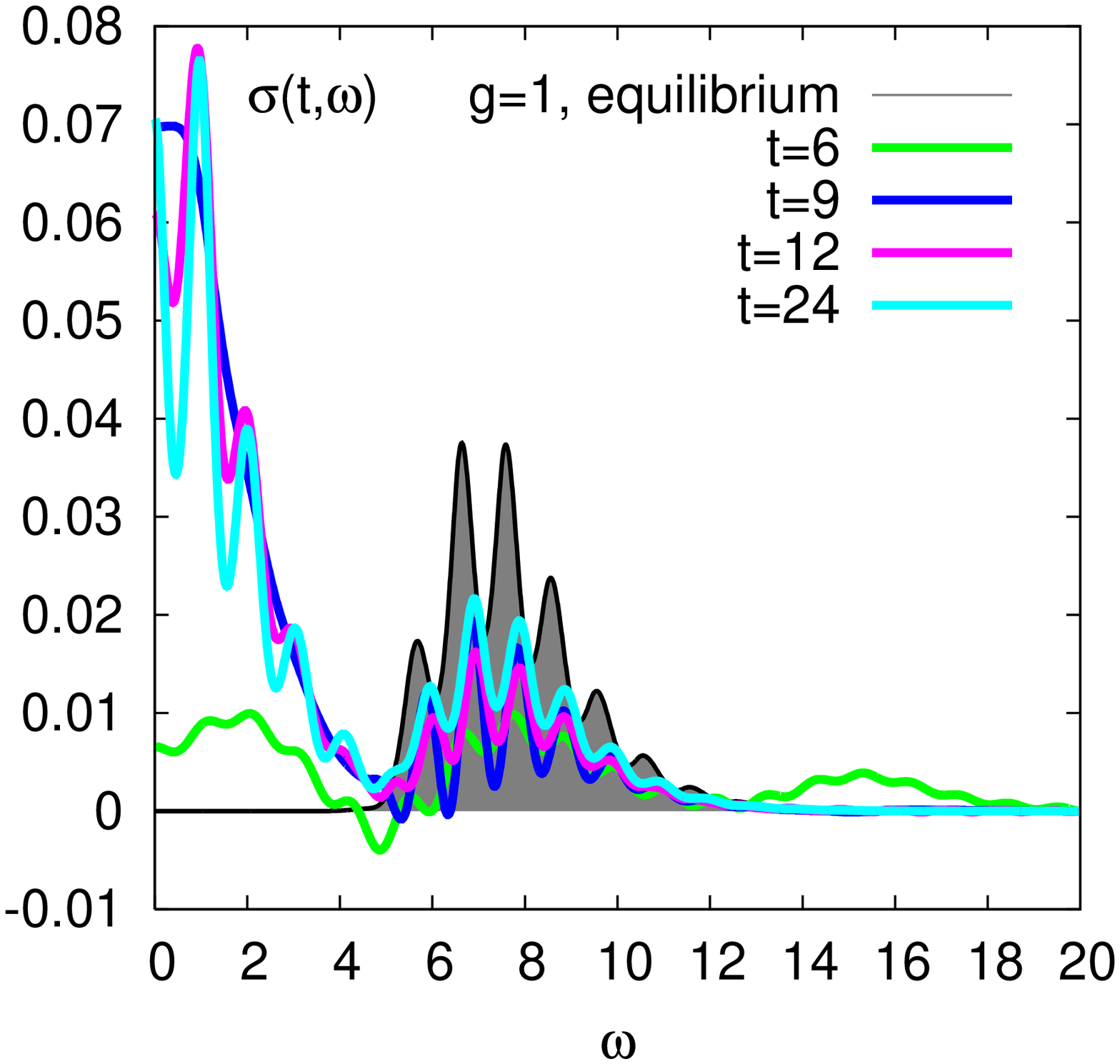}
\includegraphics[angle=-90, width=0.48\columnwidth]{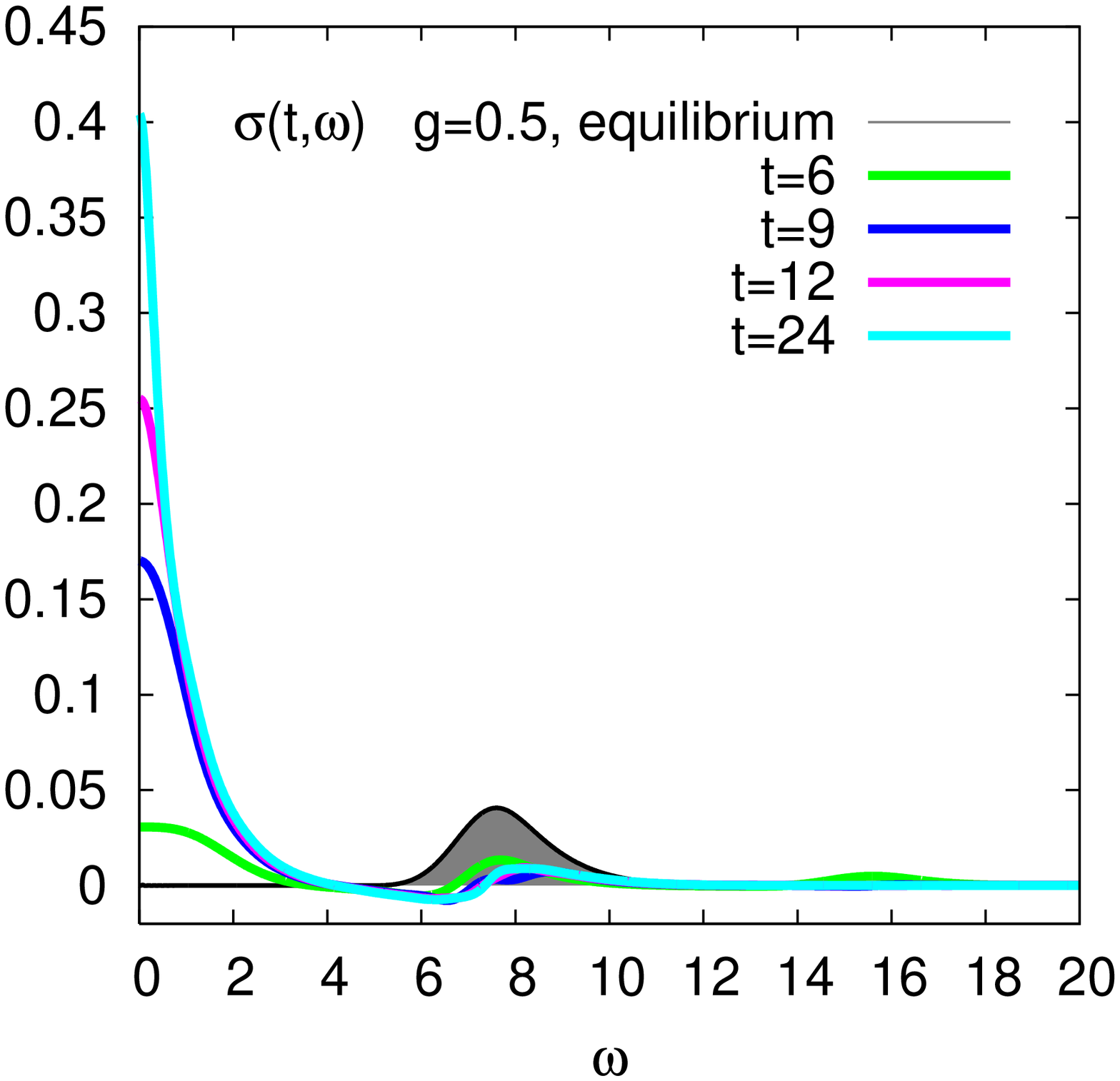}
\caption{Time-resolved optical conductivity after a pulse excitation. Left panel: $U=8, g=1$. Pulse form as in Fig.~4, but with amplitude $8$.
The doublon number $d-d_\text{eq}$ for $t=9,12,14$ is $0.16, 0.15, 0.11$ ($g=1$) and $0.21,0.22,0.21$ ($g=0.5$).
}
\label{fig_conductivity}
\end{center}
\end{figure}   

The formation of a metallic state with polaronic carriers after a photo-doping pulse is also evident in the time-resolved optical conductivity. This is not surprising, because within DMFT, the optical conductivity $\sigma(\omega,t)$ (whose definition is given in Ref.~\onlinecite{Eckstein2008}) is obtained from the convolution of two time-dependent spectral functions. Figure~\ref{fig_conductivity} shows $\sigma(\omega,t)$ for the pulse-excited system with $U=8$, $g=1$ and $g=0.5$. The pulse shape is the same as shown in Fig.~4 of the main text, but with a smaller amplitude. In the initial equilibrium state ($\beta=5$), the optical conductivity has weight around $\omega=8$, corresponding to transitions between the Hubbard bands, which for the larger phonon-coupling are split into phonon side-bands. At $t=6$, during the application of the pulse with frequency $\Omega_\text{pulse}\approx 8$, the conductivity in the low-energy region grows, and we notice some weight also at energy $\omega\approx 16$ (for simplicity, we have neglected vertex corrections, which may be relevant in the presence of an external field). The explanation for this high-energy feature is that in the periodically driven system, ``Floquet side-bands" appear, which are separated from the Hubbard bands by multiples of the driving frequency $\Omega_\text{pulse}$ \cite{Tsuji2008}. At $t=9$ the pulse is over and this high-energy feature is gone. Instead, a Drude peak appears at low frequencies. Because of the cooling effect, the peak in the model with weak phonon-coupling ($g=0.5$) grows and sharpens on a time-scale  of about 20/W, indicating the formation of a metallic state with large kinetic energy. In the model with $g=1$, the growth of the Drude peak stops already at $t\approx 9$, and its weight is considerably smaller than in the weakly coupled case (the kinetic energy of the polarons is lower). At somewhat later times periodic modulations appear in the Drude peak on the scale $\omega_0$. They arise from transitions between the partially filled polaronic sidebands. 

Since the polaronic features are clearly visible in the time-resolved optical conductivity, we expect that 
the field-induced enhancement of the electron-phonon coupling, as evidenced for example by the proposed sub-gap THz pump experiment (Fig.~5 in the main text), 
can also be seen in optical conductivity data.

\begin{figure}[hb]
\begin{center}
\includegraphics[angle=-90, width=0.48\columnwidth]{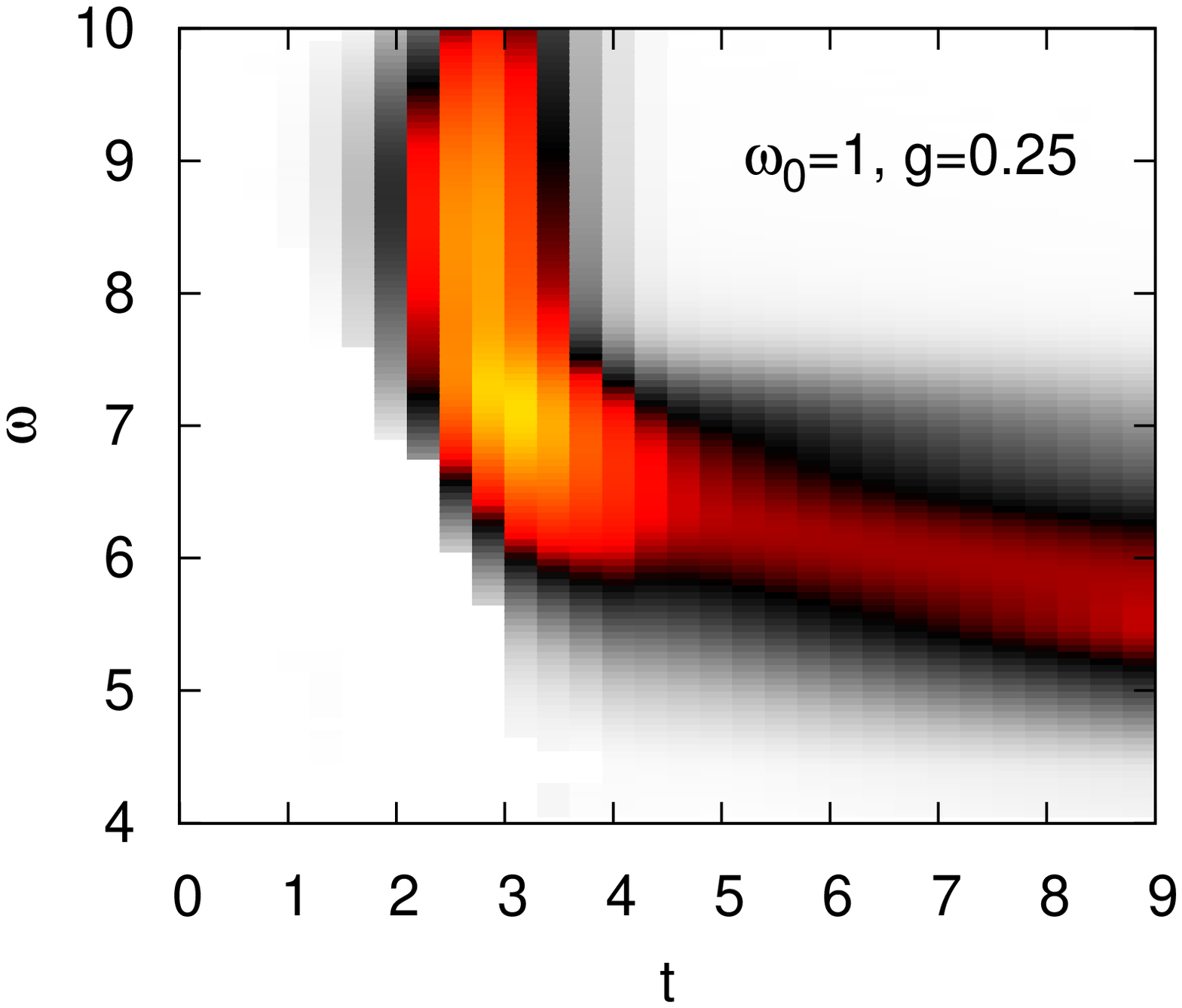}
\includegraphics[angle=-90, width=0.48\columnwidth]{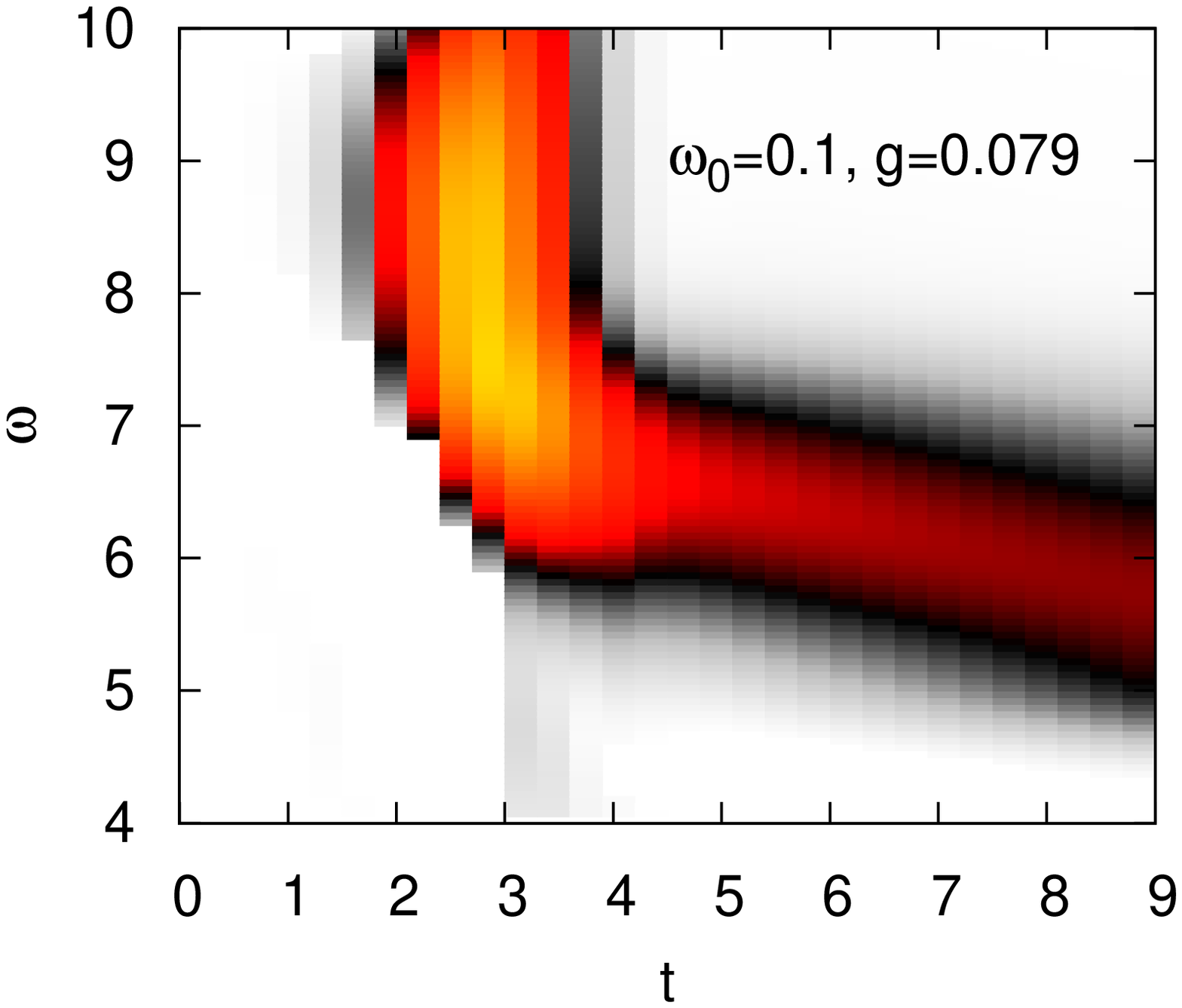}\\
\vspace{3mm}
\includegraphics[angle=-90, width=0.48\columnwidth]{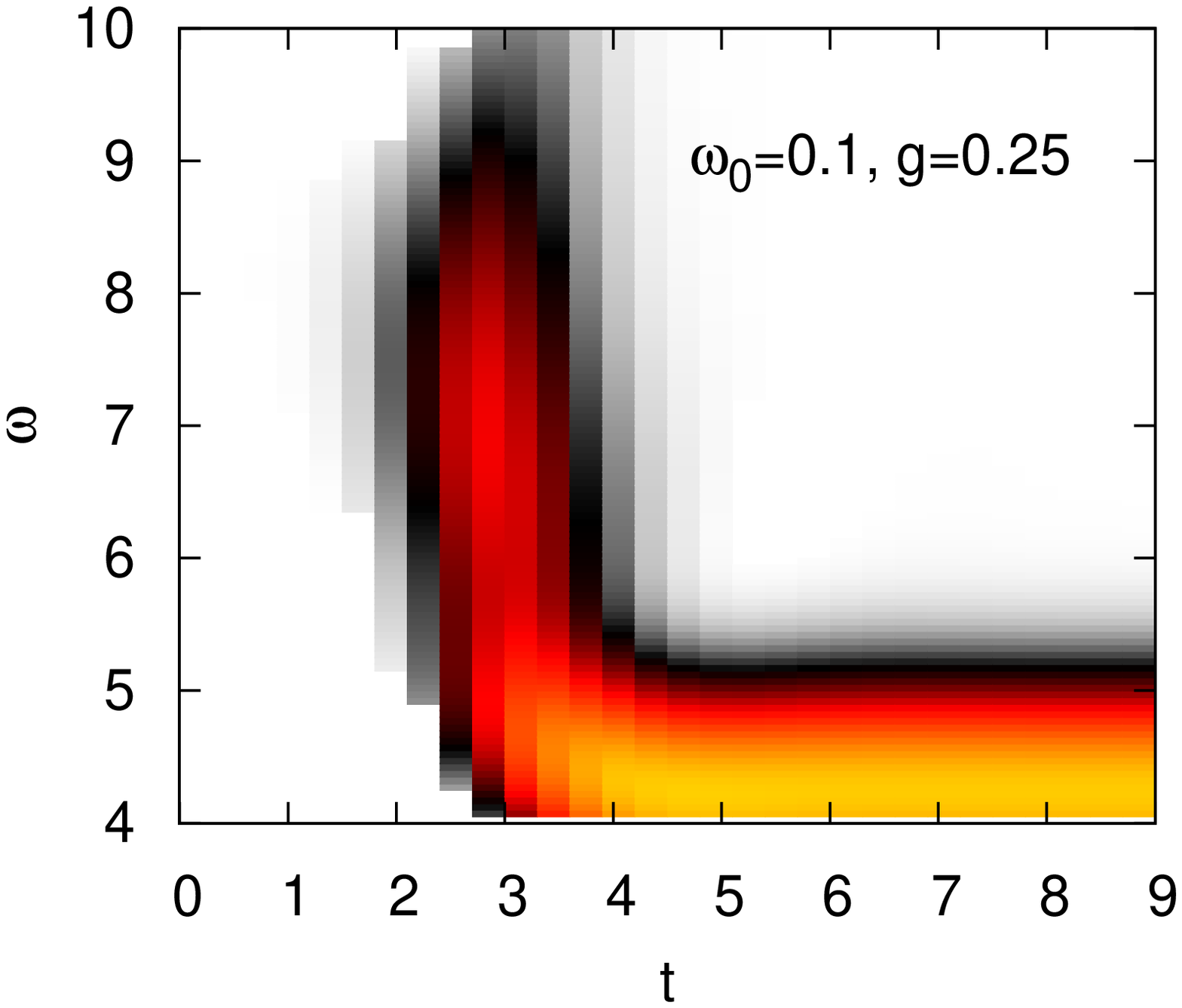}
\includegraphics[angle=-90, width=0.48\columnwidth]{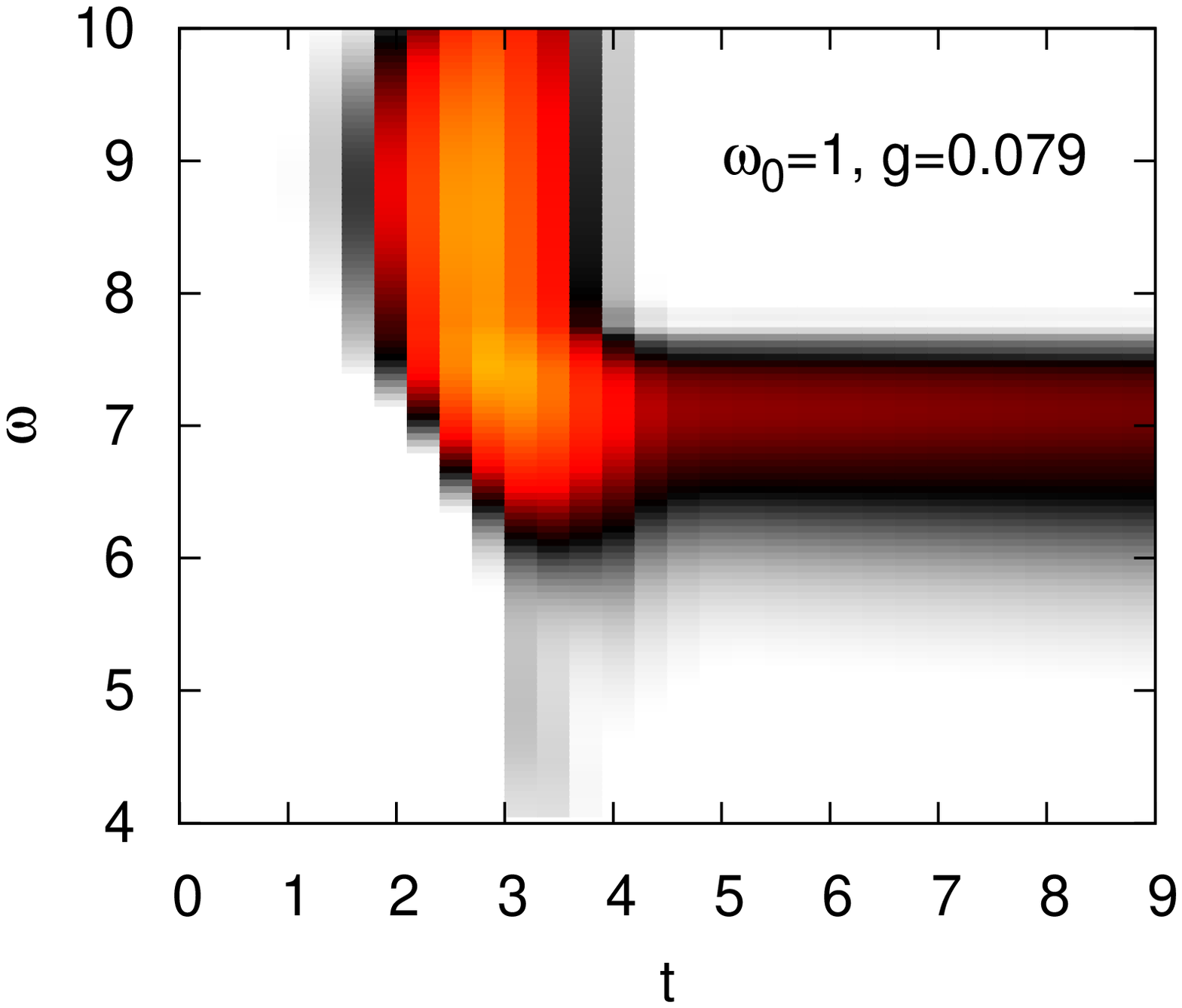}\\
\vspace{8mm}
\includegraphics[angle=-90, width=0.49\columnwidth]{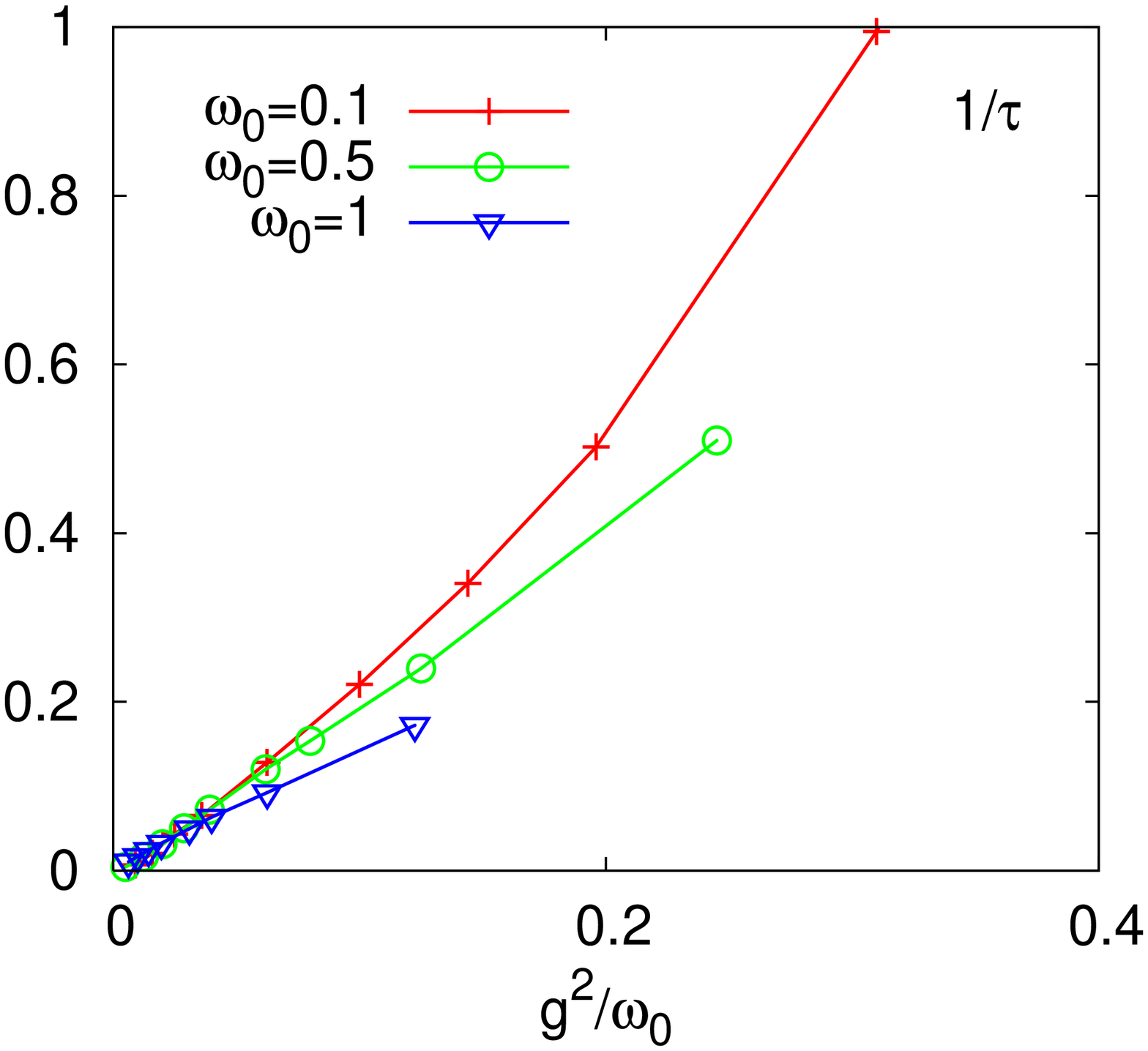}
\includegraphics[angle=-90, width=0.49\columnwidth]{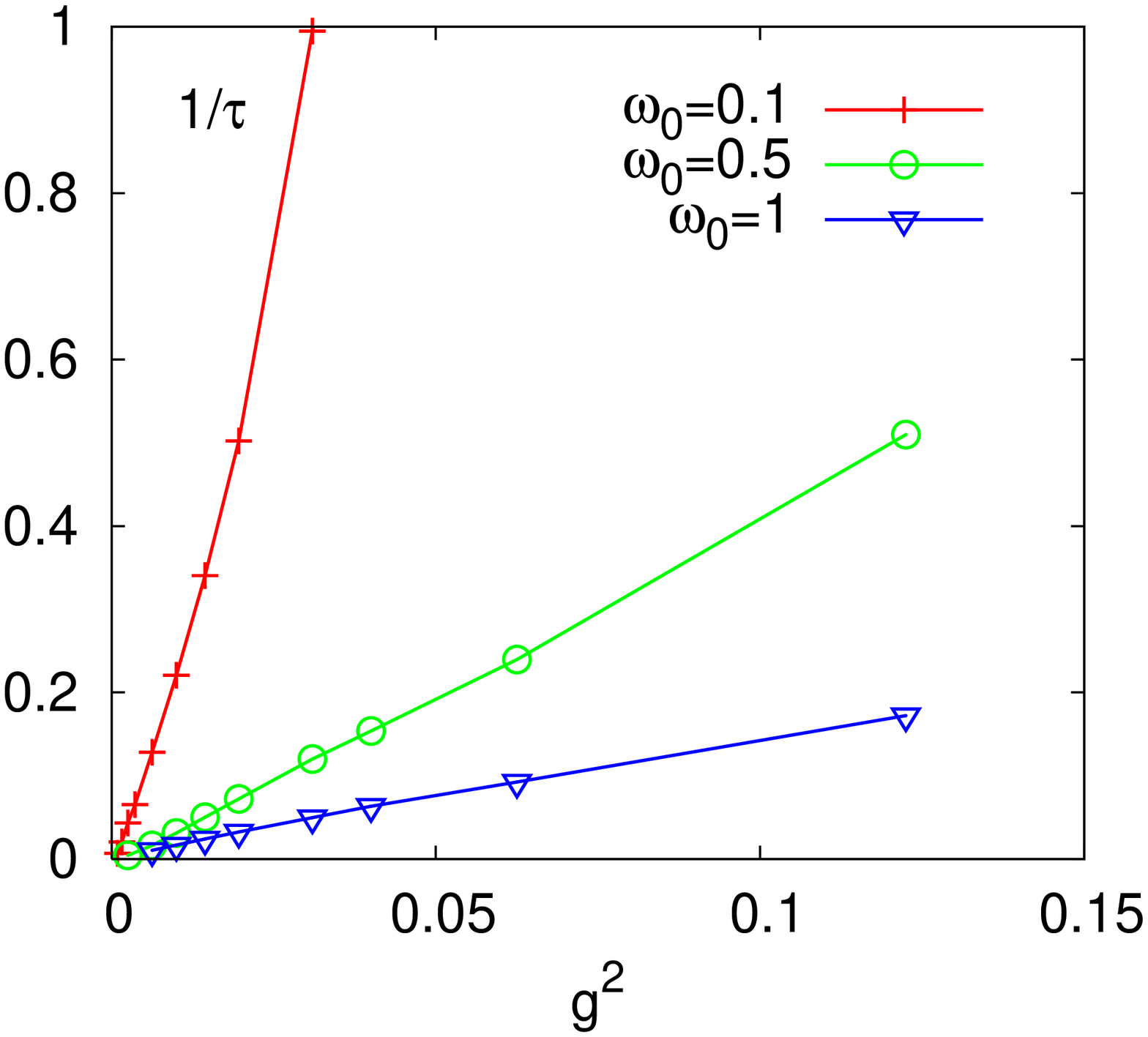} 
\caption{
Top four panels: Occupied spectral weight $A^<(t,\omega)$ in the upper Hubbard band of a photo-doped system with $U=12$ and indicated values of $g$ and $\omega_0$. The parameters in the top two panels correspond to the same $g^2/\omega_0$.
Bottom panels: relaxation time $\tau$ extracted from the kinetic energy plotted as a function of $g^2$ (right panel) and $g^2/\omega_0$ (left panel).  
}
\label{fig_cooling_omega}
\end{center}
\end{figure}

\noindent\\
{\bf Effective dissipation strength}

The energy distribution of photodoped carriers in the (paramagnetic) Hubbard model changes only very slowly, so that on the time-scales accessible with our simulations, almost no relaxation of the spectral function can be observed \cite{Eckstein2011pump}. If the electrons are coupled to phonons, the relaxation within the Hubbard band can be very fast (Fig.~5 of the main text). Electrons which are inserted at the upper edge of the Hubbard band by a photo-doping pulse dissipate energy 
to the ``heat-bath'' provided by the phonons, 
and the spectral weight drifts down in energy until it accumulates at the lower band edge. 
The corresponding drift velocity can be used as a measure for the dissipation strength in the system.  

Our numerical simulations show that neither $\omega_0$, nor the coupling strength $g$ individually determine this dissipation strength, but the ratio $g^2/\omega_0$. The top two panels of Fig.~\ref{fig_cooling_omega} plot $A^<(\omega,t)$ in two systems with $U=12$, but very different phonon frequencies and phonon coupling strengths. The parameters, $\omega_0=1$, $g=0.25$ and $\omega_0=0.1$, $g=0.079$ however correspond to the same value $g^2/\omega_0$. While the density of absorbed doublons is different in the two simulations, the drift velocity of the spectral weight is almost identical. In contrast, the model with $\omega_0=0.1, g=0.25$ ($\omega_0=1$, $g=0.079$) shows a much faster (slower) relaxation, see middle panels.  

From the approximately exponential long-time relaxation of the kinetic energy we can furthermore extract relaxation times $\tau$. The bottom right panel plots $\frac{1}{\tau}$ as a function of $g^2$ for different $\omega_0$. While for fixed $\omega_0$, the inverse relaxation time grows approximately proportional to $g^2$, it depends strongly on the value of the phonon frequency. A rough data collapse is obtained if we plot $\frac{1}{\tau}$ as a function of $g^2/\omega_0$. 
These results show that the dissipation strength is controlled by $\lambda=g^2/\omega_0$, at least if $\lambda$ is small.  


\begin{thebibliography}{99}
\bibitem{footnote_deq} We use the double occupancy in the initial equilibrium state as an estimate for $d_\text{eq}$.
\end{thebibliography}
\end{document}